\documentclass[prd,amsmath,amssymb,twocolumn,superscriptaddress,nofootinbib,nobibnotes,longbibliography]{revtex4-2}

\usepackage{xcolor,color}
\usepackage{comment}
\usepackage{graphicx}
\usepackage{url}
\usepackage{mathtools}
\usepackage{makecell}
\usepackage{dcolumn}
\usepackage{bm}
\usepackage{xspace}
\usepackage{multirow}

\usepackage[colorlinks = true,
            linkcolor = red,
            urlcolor  = blue,
            citecolor = red,
            anchorcolor = blue]{hyperref}

\newcommand{\Eb}{$E_\text{blip}$\xspace} 
 
\newcommand{\MeVee}{MeV$_{\text{ee}}$\xspace}

\usepackage{lineno}

\begin{document}

\title{Demonstration of new MeV-scale capabilities in large neutrino LArTPCs using ambient radiogenic and cosmogenic activity in MicroBooNE}

\newcommand{\ANL}{Argonne National Laboratory (ANL), Lemont, IL, 60439, USA}
\newcommand{\Bern}{Universit{\"a}t Bern, Bern CH-3012, Switzerland}
\newcommand{\BNL}{Brookhaven National Laboratory (BNL), Upton, NY, 11973, USA}
\newcommand{\UCSB}{University of California, Santa Barbara, CA, 93106, USA}
\newcommand{\Cambridge}{University of Cambridge, Cambridge CB3 0HE, United Kingdom}
\newcommand{\CIEMAT}{Centro de Investigaciones Energ\'{e}ticas, Medioambientales y Tecnol\'{o}gicas (CIEMAT), Madrid E-28040, Spain}
\newcommand{\Chicago}{University of Chicago, Chicago, IL, 60637, USA}
\newcommand{\Cincinnati}{University of Cincinnati, Cincinnati, OH, 45221, USA}
\newcommand{\CSU}{Colorado State University, Fort Collins, CO, 80523, USA}
\newcommand{\Columbia}{Columbia University, New York, NY, 10027, USA}
\newcommand{\Edinburgh}{University of Edinburgh, Edinburgh EH9 3FD, United Kingdom}
\newcommand{\FNAL}{Fermi National Accelerator Laboratory (FNAL), Batavia, IL 60510, USA}
\newcommand{\Granada}{Universidad de Granada, Granada E-18071, Spain}
\newcommand{\IIT}{Illinois Institute of Technology (IIT), Chicago, IL 60616, USA}
\newcommand{\ICL}{Imperial College London, London SW7 2AZ, United Kingdom}
\newcommand{\Indiana}{Indiana University, Bloomington, IN 47405, USA}
\newcommand{\KSU}{Kansas State University (KSU), Manhattan, KS, 66506, USA}
\newcommand{\Lancaster}{Lancaster University, Lancaster LA1 4YW, United Kingdom}
\newcommand{\LANL}{Los Alamos National Laboratory (LANL), Los Alamos, NM, 87545, USA}
\newcommand{\Louisiana}{Louisiana State University, Baton Rouge, LA, 70803, USA}
\newcommand{\Manchester}{The University of Manchester, Manchester M13 9PL, United Kingdom}
\newcommand{\MIT}{Massachusetts Institute of Technology (MIT), Cambridge, MA, 02139, USA}
\newcommand{\Michigan}{University of Michigan, Ann Arbor, MI, 48109, USA}
\newcommand{\MSU}{Michigan State University, East Lansing, MI 48824, USA}
\newcommand{\Minnesota}{University of Minnesota, Minneapolis, MN, 55455, USA}
\newcommand{\Nankai}{Nankai University, Nankai District, Tianjin 300071, China}
\newcommand{\NMSU}{New Mexico State University (NMSU), Las Cruces, NM, 88003, USA}
\newcommand{\Oxford}{University of Oxford, Oxford OX1 3RH, United Kingdom}
\newcommand{\Pitt}{University of Pittsburgh, Pittsburgh, PA, 15260, USA}
\newcommand{\QMUL}{Queen Mary University of London, London E1 4NS, United Kingdom}
\newcommand{\Rutgers}{Rutgers University, Piscataway, NJ, 08854, USA}
\newcommand{\SLAC}{SLAC National Accelerator Laboratory, Menlo Park, CA, 94025, USA}
\newcommand{\SDSMT}{South Dakota School of Mines and Technology (SDSMT), Rapid City, SD, 57701, USA}
\newcommand{\Maine}{University of Southern Maine, Portland, ME, 04104, USA}
\newcommand{\Syracuse}{Syracuse University, Syracuse, NY, 13244, USA}
\newcommand{\TelAviv}{Tel Aviv University, Tel Aviv, Israel, 69978}
\newcommand{\UTA}{University of Texas, Arlington, TX, 76019, USA}
\newcommand{\Tufts}{Tufts University, Medford, MA, 02155, USA}
\newcommand{\VTech}{Center for Neutrino Physics, Virginia Tech, Blacksburg, VA, 24061, USA}
\newcommand{\Warwick}{University of Warwick, Coventry CV4 7AL, United Kingdom}

\affiliation{\ANL}
\affiliation{\Bern}
\affiliation{\BNL}
\affiliation{\UCSB}
\affiliation{\Cambridge}
\affiliation{\CIEMAT}
\affiliation{\Chicago}
\affiliation{\Cincinnati}
\affiliation{\CSU}
\affiliation{\Columbia}
\affiliation{\Edinburgh}
\affiliation{\FNAL}
\affiliation{\Granada}
\affiliation{\IIT}
\affiliation{\ICL}
\affiliation{\Indiana}
\affiliation{\KSU}
\affiliation{\Lancaster}
\affiliation{\LANL}
\affiliation{\Louisiana}
\affiliation{\Manchester}
\affiliation{\MIT}
\affiliation{\Michigan}
\affiliation{\MSU}
\affiliation{\Minnesota}
\affiliation{\Nankai}
\affiliation{\NMSU}
\affiliation{\Oxford}
\affiliation{\Pitt}
\affiliation{\QMUL}
\affiliation{\Rutgers}
\affiliation{\SLAC}
\affiliation{\SDSMT}
\affiliation{\Maine}
\affiliation{\Syracuse}
\affiliation{\TelAviv}
\affiliation{\UTA}
\affiliation{\Tufts}
\affiliation{\VTech}
\affiliation{\Warwick}

\author{P.~Abratenko} \affiliation{\Tufts}
\author{O.~Alterkait} \affiliation{\Tufts}
\author{D.~Andrade~Aldana} \affiliation{\IIT}
\author{L.~Arellano} \affiliation{\Manchester}
\author{J.~Asaadi} \affiliation{\UTA}
\author{A.~Ashkenazi}\affiliation{\TelAviv}
\author{S.~Balasubramanian}\affiliation{\FNAL}
\author{B.~Baller} \affiliation{\FNAL}
\author{A.~Barnard} \affiliation{\Oxford}
\author{G.~Barr} \affiliation{\Oxford}
% ...

\collaboration{The MicroBooNE Collaboration}
\thanks{microboone\_info@fnal.gov}\noaffiliation


\date{\today}

\begin{abstract}
Large neutrino liquid argon time projection chamber (LArTPC) experiments can broaden their physics reach by reconstructing and interpreting  MeV-scale energy depositions, or blips, present in their data. We demonstrate new calorimetric and particle discrimination capabilities at the MeV energy scale using reconstructed blips in data from the MicroBooNE LArTPC at Fermilab. 
We observe a concentration of low energy ($<$3~MeV) blips around fiberglass mechanical support struts along the TPC edges with energy spectrum features consistent with the Compton edge of 2.614 MeV $^{208}$Tl decay $\gamma$~rays. These features are used to verify proper calibration of electron energy scales in MicroBooNE's data to few percent precision and to measure the specific activity of $^{208}$Tl in the fiberglass composing these struts, $(11.7 \pm 0.2 ~\text{(stat)}  \pm  3.1~\text{(syst)})$~Bq/kg. Cosmogenically-produced blips above 3~MeV in reconstructed energy are used to showcase the ability of large LArTPCs to distinguish between low-energy proton and electron energy depositions. An enriched sample of low-energy protons selected using this new particle discrimination technique is found to be smaller in data than in dedicated CORSIKA cosmic ray simulations, suggesting either incorrect CORSIKA modeling of incident cosmic fluxes or particle transport modeling issues in Geant4.  
 
\end{abstract}

\maketitle

\section{Introduction} \label{sec:introduction}

Liquid argon time projection chambers (LArTPCs) are unique among existing deployed neutrino detector technologies in combining millimeter-scale position resolution with sub-MeV energy detection thresholds. 
In a LArTPC, MeV-scale and sub-MeV charged particles generate ionization electron clouds with $\mu$m to cm spatial extents, which efficiently drift through the liquid argon environment in a uniform electric field towards planes of closely-spaced conducting wires.  
After processing the wire signals, these electron clouds appear as isolated compact features, or `blips', in a LArTPC event, contrasting the extended shower-like or track-like features produced by higher-energy charged particles.  
The position resolution of neutrino LArTPCs is defined by the spacing of these sense wires (3~to~5~mm in current  LArTPCs~\cite{Anderson:2012vc,ub_det,lariat_detpaper,ICARUS:2023gpo}), while detection thresholds are defined by the noise levels achieved by sense wire readout electronics (as low as 300 $e^-$ in equivalent noise charge per ADC sample in the MicroBooNE LArTPC, for example~\cite{MicroBooNE:2017qiu}).  

The combination of precise position resolution and low energy thresholds can be exploited to expand the scope of particle physics that can be performed by GeV-scale accelerator neutrino LArTPC experiments.  
Existing neutrino LArTPCs can use these capabilities to identify neutrino-argon ($\nu$-Ar) interactions with important yet under-studied  attributes.  
For example, $\nu$-Ar final-state neutrons, visible primarily as proton- or electron-induced low-energy activity~\cite{argo_mev, MicroBooNE:2024hun, Rivera:2021dcf}, will play an important role in defining energy reconstruction biases for neutrinos and antineutrinos in future leptonic CP-violation measurements~\cite{Ankowski:2015kya,Friedland:2018vry,DUNE:2020txw}.  
Low-activity $\nu$-Ar vertices, such as those generated by low momentum-transfer neutral-current scatters, can be overlooked by standard reconstruction tools despite their potential value in probing the nature of the long-standing MiniBooNE neutrino experiment anomaly~\cite{MiniBooNE:2020pnu,MicroBooNE:2021zai,MicroBooNE:2021nxr,MicroBooNE:2021wad}.  
More generally, the definition of `zero-proton' ($0p$) and `one or more protons' (N$p$) neutrino final-state topologies in recent LArTPC studies~\cite{MicroBooNE:2021zai,MicroBooNE:2021nxr,MicroBooNE:2021wad,MicroBooNE:2022tdd,MicroBooNE:2022zhr,MicroBooNE:2024xod,MicroBooNE:2024klj,MicroBooNE:2024tmp} is based on a LArTPC's effective threshold for proton reconstruction (20 or 35~MeV on kinetic energy, depending on the analysis).   However, low-energy reconstruction tools may enable reliable reconstruction of proton activity below these thresholds, increasing precision in knowledge of the hadronic content of final-state $\nu$-Ar interactions.  
MeV-scale activity can also enable searches for low-energy muon or meson decay-at-rest neutrino interactions~\cite{BenevidesRodrigues:2022wxz,Grant:2015jva,Harnik:2019iwv} or new physics beyond the Standard Model (BSM)~\cite{Castiglioni:2020tsu,sbnd_phys,leplar_paper}. 
In the coming decade, application of low-energy LArTPC capabilities in the Deep Underground Neutrino Experiment (DUNE) far detector will be crucial to calibrating detector response~\cite{DUNE:2020txw} and to achieving sensitivity to supernova and solar neutrino $\nu$-Ar interactions~\cite{DUNE:2020zfm,Q-Pix:2022zjm,dune_solar}.  
A range of potential physics benefits are surveyed in Refs.~\cite{Castiglioni:2020tsu,leplar_paper}.  

The exploration of the MeV-scale physics capabilities of neutrino LArTPCs is a relatively recent effort.  
Following early studies of $\sim$5-50~MeV cosmic muon decay electrons in the ICARUS and MicroBooNE detectors~\cite{icarus_michel,ub_michel}, data from the ArgoNeuT LArTPC was used to detect isolated blips produced by final-state photons and neutrons from $\nu$-Ar interactions~\cite{argo_mev}, and to search for similar features from hypothetical beam-produced millicharged particles~\cite{argo_mcp}. The MeV-scale capabilities of the MicroBooNE detector were first studied in \cite{ub_mev,ub_ar39,ub_mev_pub_note}. The MicroBooNE LArTPC was used to measure the properties of decay radiation from $^{222}$Rn progeny in its liquid argon bulk~\cite{ub_radon}, including $^{214}$Po decay $\alpha$-particles generating $\sim$~100-200~keV worth of electron-equivalent ionization~\cite{MicroBooNE:2023ftv}.  
That analysis also used $^{214}$Bi decay electrons to demonstrate other MeV-scale calorimetric capabilities, showing data-simulation energy scale agreement at $<$5\% precision and a modeled energy resolution better than 10\% above 1~MeV.  
A first study of blip-based particle discrimination was recently performed by LArIAT using purified samples of stopping $\mu^-$ and $\pi^-$~\cite{LArIAT:2024otd}.  
With one exception~\cite{lariat_michels}, MeV-scale measurements to date have focused on low-energy capabilities of neutrino LArTPC ionization charge collection systems, bypassing consideration of associated scintillation light signals.

In this study, we extend the range of demonstrated neutrino LArTPC capabilities and tools available at the MeV scale using ambient radiogenic and cosmogenic blips reconstructed in MicroBooNE data. Leveraging an intrinsic detector source of $^{208}$Tl decay $\gamma$ rays, we perform a percent-level precision energy scale calibration for reconstructed blips, providing a template for blip-based energy scale calibrations in DUNE and other future LArTPC experiments.  
A byproduct of this calibration is a measurement of the specific activity of $^{208}$Tl in MicroBooNE's fiberglass TPC support struts that highlights the importance of radio-purity screening in DUNE and other next-generation LArTPCs. 

Next, we develop a new particle identification (PID) metric for blips based on size-to-energy comparisons, capable of distinguishing low-energy charged particles of differing stopping powers in argon, such as electrons and protons.  
We demonstrate the value of this PID capability by identifying, for the first time in a neutrino LArTPC, a sub-dominant population of isolated cosmogenic protons in MicroBooNE. 
Comparisons to CORSIKA cosmogenic simulations reveal an excess of measured protons in CORSIKA, indicating potential cosmogenic flux or cosmic ray particle transport issues in the CORSIKA or Geant4 toolkits used by MicroBooNE. 

Section~\ref{sec:microboone} summarizes the MicroBooNE detector, while Sec.~\ref{sec:reconstruction} describes the analysis tools used to process MicroBooNE data and reconstruct MeV-scale blips.  
Datasets and Monte Carlo (MC) simulations used in this analysis are described in Sec.~\ref{sec:datasets}.  
After a general description of all selected blips in Sec.~\ref{sec:analysis}, Sec.~\ref{sec:hotspot} provides a detailed analysis of the $^{208}$Tl-produced 2.614 MeV $\gamma$-ray Compton edge.  
Section~\ref{sec:cosmic} describes the developed PID metric and cosmogenic proton analysis, with a summary of study outcomes provided in Sec.~\ref{sec:summary}.

\section{The MicroBooNE Detector} \label{sec:microboone}

MicroBooNE was a single-phase LArTPC detector located in the Booster Neutrino Beamline (BNB) at Fermi National Accelerator Laboratory that operated from 2015 to 2021.  
The TPC, its cryostat, and all supporting detector infrastructure were located in the Liquid Argon Test Facility, an on-surface building providing minimal overburden shielding the LArTPC from cosmic rays.  
While details of the MicroBooNE detector and support systems are presented in Ref.~\cite{ub_det}, we will briefly overview MicroBooNE design elements relevant to this work.  

\begin{figure}
\includegraphics[width=0.47\textwidth]{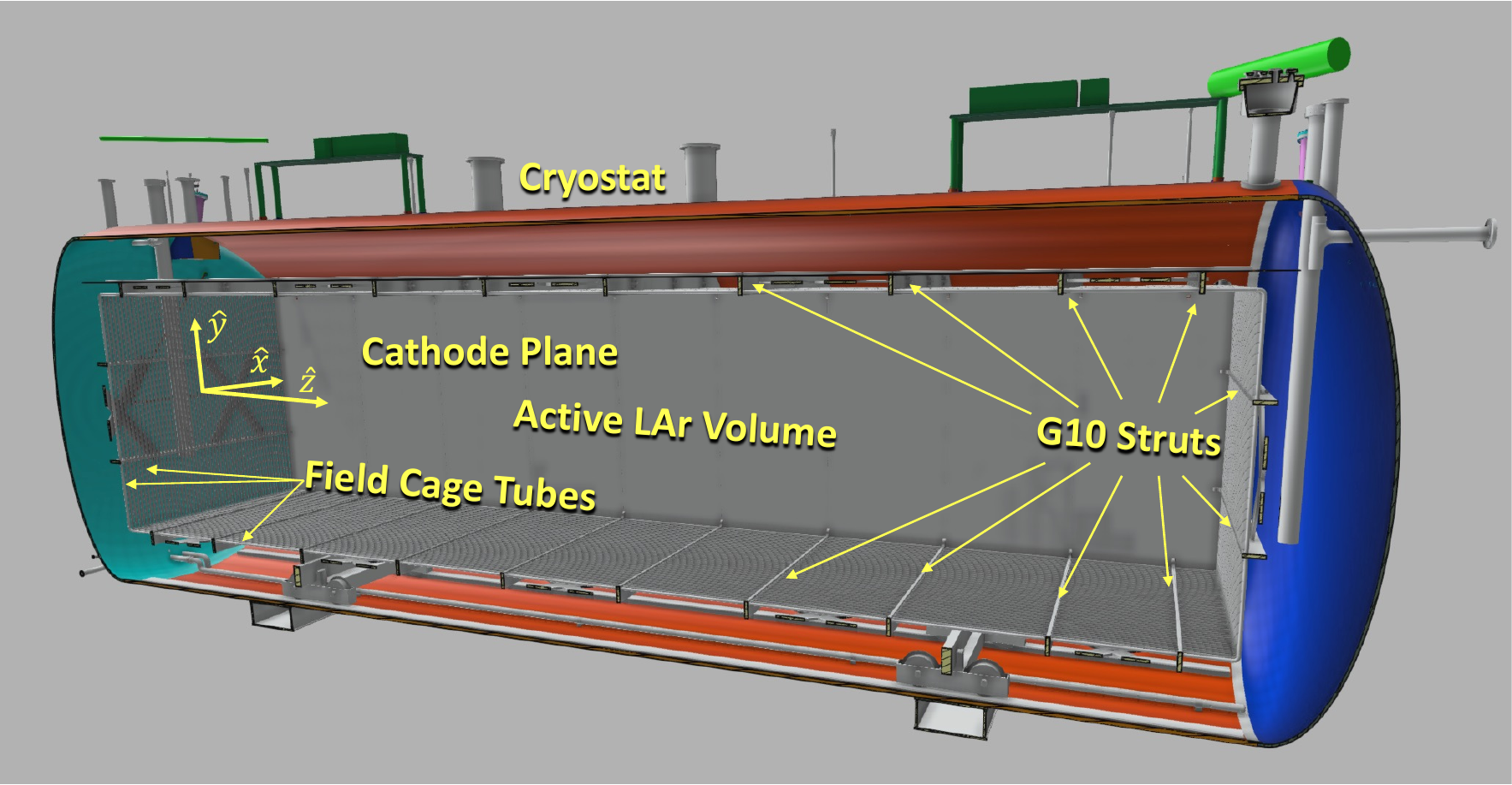}
\caption{A labeled isometric model of MicroBooNE's cryostat and TPC interior. The model shows the TPC viewed from its anode plane wire side. The coordinate system $(\hat{x},\hat{y},\hat{z})$ is indicated, with the neutrino beam oriented along the $+z$ direction.}
\label{fig:TPCdrawing}
\end{figure}

The primary component of the MicroBooNE detector was a $2.56 \text{ m width}\times 2.33 \text{ m height}\times 10.37 \text{ m length}$ TPC containing 85~metric tons of purified LAr.  
As illustrated in Fig.~\ref{fig:TPCdrawing}, the TPC formed a rectangular prism, with a uniform 274\,V/cm electric field between the two vertically-oriented long faces of the TPC.  
This electric field, oriented in the $x$ direction (indicated in Fig.~\ref{fig:TPCdrawing}) in MicroBooNE's coordinate system, was generated by a stainless steel cathode plane charged to -70~kV at $x=256$~cm and three anode planes of conducting readout wires at $x\approx0$~cm, spaced 3~mm apart, with a 3~mm separation gap between planes. 
A series of 64 stainless steel field cage tubes of 2.54~cm diameter surrounded the active LAr volume, running perpendicular to the electric field along the walls of the TPC. 
A series of voltage divider circuits were used to step the field down linearly along the field cage from the cathode to the anode to ensure a uniform electric field within the active volume.

\begin{figure}[h]
\includegraphics[width=0.47\textwidth]{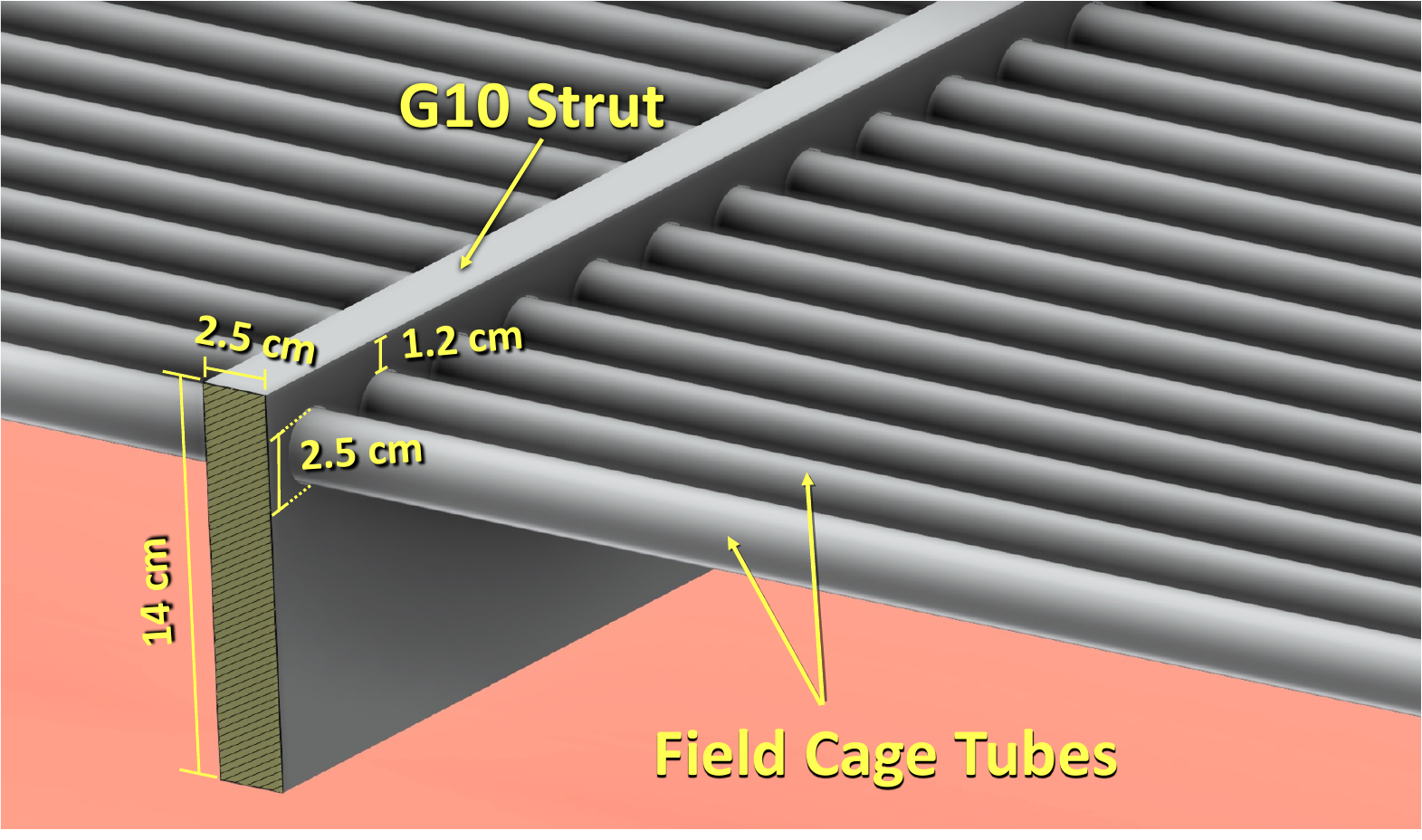}
\caption{A close-up labeled isometric drawing of the interface between the bottom TPC field cage and support strut elements.  The pictured G10 strut provides mechanical support to the TPC while maintaining electrical insulation of all field cage elements.}
\label{fig:TPCdrawing2}
\end{figure}

The cathode and anode surfaces of the TPC were mechanically supported by metal structures, while field cage tubes were stabilized using struts of 2.5~cm thick G10, an electrically insulating epoxied fiberglass laminate~\cite{G10}.  
The G10 struts also provide the only mechanical connection between anode and cathode structures, enforcing a fixed parallel orientation between the two planes.  
Ten (two) elongated 15.4~kg struts were present on each TPC top and bottom (front and back) surface, with each strut's long axis running the entire $x$ length of the TPC active volume.  
As shown in Fig.~\ref{fig:TPCdrawing2}, field cage tubes penetrated machined holes centered 2.5~cm from one side of the strut's 14~cm height in $y$, leaving 1.2~cm of G10 material extending beyond the field cage tubes.  
Thus, G10 is the only non-LAr material present in the active volume of MicroBooNE's TPC.

Charged particle interactions within the active TPC volume generate ionized electrons which drift in the uniform electric field at a speed of 1.1\,mm/$\mu$s towards anode plane readout wires.  
Electrical signals generated on the conducting wires in response to nearby ionization charge are sensed, processed, and recorded by readout electronics with a sampling period of 500 ns as described in Refs.~\cite{MicroBooNE:2017qiu,MicroBooNE:2018swd,MicroBooNE:2018vro}.  
By collecting 6400 ADC samples (3.2 ms) per wire per detector trigger, all ionization charge present inside the TPC at the time of triggering can be recorded regardless of drift distance in $x$.  

The 2,400 wires in each of the two planes closest to the active TPC volume are oriented at $\pm$~60$^{\circ}$ with respect to the vertical direction.    
These 'induction' planes are voltage biased at -110 V and 0 V to ensure maximal electron transparency, causing charge to cross these planes and induce bipolar signals on these wires.  The 3456 vertically oriented wires in the third 'collection' plane are held at +220 V, and record unipolar ADC waveform pulses as ionization charge stops and is collected at this plane. Raw wire signals are further processed and noise-filtered to obtain charge waveforms upon which higher-level reconstruction and analysis is performed~\cite{MicroBooNE:2017qiu,MicroBooNE:2018swd,MicroBooNE:2018vro}. MicroBooNE's charge collection system is sensitive to LArTPC energy depositions well below 1~MeV due to: (a) the achieved residual equivalent noise charge of 300 $e^-$ and 400 $e^-$ on the longest collection and induction wires, respectively, considered alongside the mean argon ionization energy of 23.6 eV per electron; (b) the electron-ion recombination survival fraction in MicroBooNE’s electric field of 60\% for minimally-ionizing particles; and (c) MicroBooNE’s long drift electron lifetime.

Scintillation light is also generated in a LArTPC over comparatively short (a few\,$\mu s$) timescales following ionization and excitation of the LAr by charged particles.  
While the MicroBooNE detector uses a photomultiplier-based light collection system to take advantage of scintillation signatures from higher-energy particle interactions, the magnitude of light collection in most detector regions is too low to enable sensitivity to MeV-scale signatures.  
For this reason, light is not used for the studies described in this paper.  
This results in complete ambiguity in the $x$ position of MeV-scale energy depositions within the MicroBooNE TPC, since the time of energy deposition relative to the detector trigger time, and thus the drift time, cannot be known.  

\section{Reconstructing and Selecting MeV-Scale Activity} \label{sec:reconstruction}
MeV-scale charged particles in LAr leave ionization trails with lengths on the order of millimeters.
For large LArTPCs with wire spacings on the order of MicroBooNE’s (3~mm), these ionization clouds will be sensed by only a few wires at most on each anode plane. Existing software targeted at reconstructing high-energy charged particles depositing 10s to 100s of MeV cannot properly identify or reconstruct the attributes of these low-energy particles.  
In this section, we will summarize how isolated MeV-scale energy depositions, or blips, are reconstructed in MicroBooNE, describe some important physics attributes of reconstructed blip objects, and outline the selection used to define a signal blip sample.  

\subsection{Blip Reconstruction}
\label{subsec:reco}

The reconstruction of blips in MicroBooNE is carried out using a \texttt{LArSoft}~\cite{larsoft} custom-built toolkit called \texttt{BlipReco}, validated in previous MicroBooNE results and described in detail in Ref.~\cite{MicroBooNE:2023ftv}.
Reconstructed blips are formed from TPC wire `hits' which contain information about pulses on a wire's filtered and deconvolved charge waveform. 
In this analysis, raw waveform deconvolution, noise filtering, and region-of-interest (ROI) identification is performed with the \texttt{WireCell} toolkit~\cite{MicroBooNE:2018swd,MicroBooNE:2018vro}.  
Hit reconstruction is performed by the \texttt{GausHit} algorithm~\cite{Baller:2017ugz}, which searches for threshold-crossing points and determines hit times, widths and amplitudes by applying Gaussian fits to these regions.  
All reconstructed hits in an event are then passed to \texttt{BlipReco},   
which forms hit `clusters' by grouping hits together on each plane based on their relative proximity to one another in wire and readout time. The proximity requirement for clustering scales with the width (in time) of each individual hit. Only hits from same or adjacent wires are grouped together.
This clustering proceeds iteratively until no new hits on the plane can be added to the cluster.
If an isolated hit is not accompanied by adjacent hits on its plane, it is also treated as a cluster.

\begin{figure}
\includegraphics[width=0.44\textwidth]{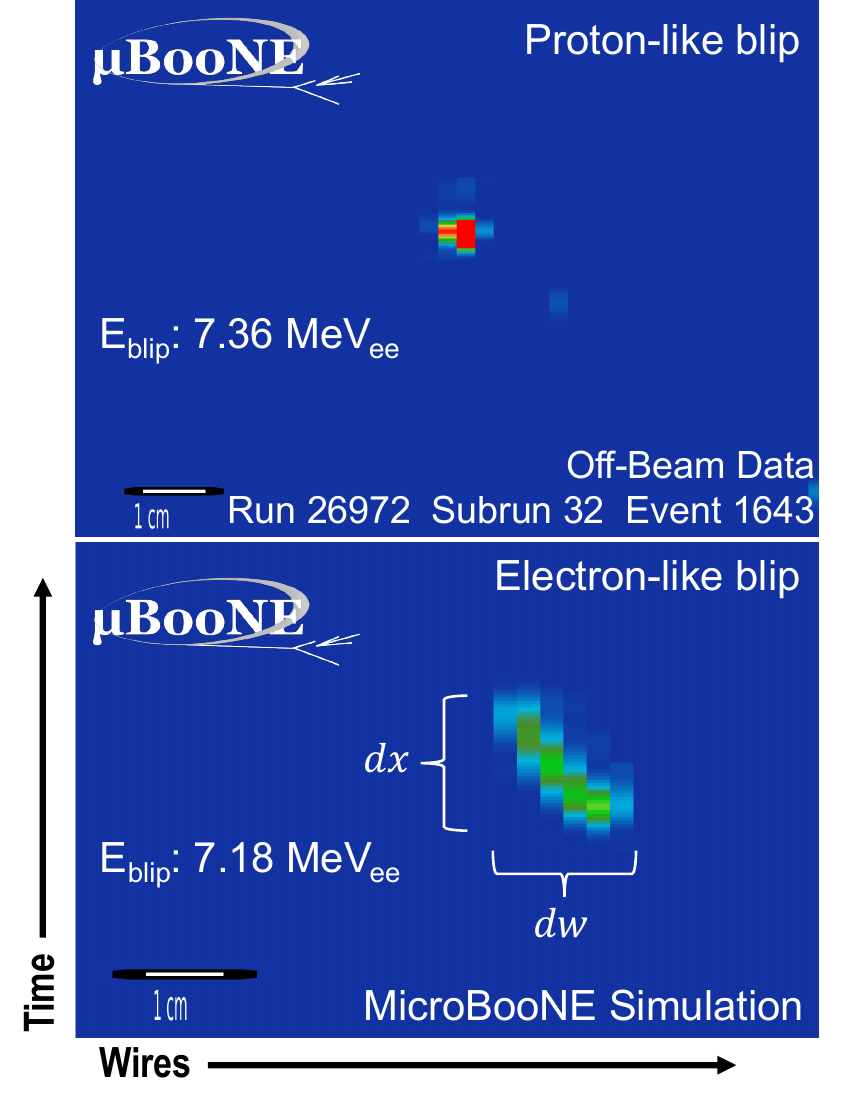}
\caption{A proton-like blip (top) from cosmic data and a low-energy electron blip (bottom) from simulation, as they appear in the MicroBooNE TPC event displays. Vertical columns represent individual wires on a wire plane, with the color scale indicating the relative charge collected at each ADC sample. This particular feature results in a cluster of hits spanning 4 (top) and 6 (bottom) wires respectively. The metrics $dx$ and $dw$, described in the text, correspond to the measured lengths of the cluster along the drift direction and projected along the axis perpendicular to wires.
}
\label{fig:blip_size}
\end{figure}

Since MeV-scale depositions span lengths approaching the inherent limiting spatial resolution of the TPC, reconstructing their true physical extent is difficult. For this analysis, the length of the cluster projected onto the axis perpendicular to the orientation of the wires on the plane, termed $dw$, is determined by multiplying the number of wires by the wire spacing of 3~mm. Similarly, the cluster's extent along the drift direction, $dx$, is calculated using the overall time span of the cluster multiplied by the electron drift velocity in LAr. The full time span is calculated as $dt = \big[ (t_1+\text{RMS}_1) - (t_0-\text{RMS}_0) \big]$, where $t_{0}$ $(t_1)$ and $\text{RMS}_{0}$ $(\text{RMS}_1)$ denote the hit time and Gaussian-fitted width of the earliest (latest) hit in the cluster. These two metrics $dx$ and $dw$, depicted visually in Fig.~\ref{fig:blip_size}, allow us to approximate the physical geometric extent of the contiguous trail of ionization that formed the cluster. 

Clusters generated on different planes by the same drifting electron cloud are then matched between planes to form three-dimensional blip objects.
The plane-matching procedure incorporates information about how closely the charge-weighted mean time of the clusters coincide, the overlap of their time spans, consistency in cluster charges, and physical wire crossing locations.  
While a blip is allowed to contain matched clusters from two or three planes, it must contain a collection plane cluster since this is the plane used for calorimetry.  

Plane-matched blips are assigned location, energy, and size metrics based on the attributes of its clusters. 
The total size ($ds$) or extent of the blip in 3D space is determined by adding in quadrature $dx$ from the collection plane cluster together with the largest $dw$ from any of the matched planes, which represents an estimate of the blip's projected length onto the anode ($y$-$z$) plane. Three-dimensional blip location, determined from the wire crossing point and the projected distance along the drift axis, is given in the same $(x,y,z)$ coordinates assigned to other reconstructed objects like tracks and showers.  
Length and location metrics in the $y$-$z$ coordinate plane have resolutions limited by the 3~mm wire spacing.

Reconstructed blip energy, $E_\text{blip}$,   is defined based on the number of electrons $Q$ reconstructed on the collection plane, the mean argon ionization energy $W_\text{ion}=23.6~\text{eV}$ \cite{PhysRevA.9.1438}, and a linear charge-energy conversion based on recombination levels expected for electrons in this energy regime, 
\begin{equation} \label{eq:electronEquivEnergy}
     E_\text{blip} [\text{MeV}_\text{ee}] = \frac{Q}{0.584} \times W_\text{ion}.  
\end{equation}

As described in detail in Ref.~\cite{MicroBooNE:2023ftv}, this linear charge-energy conversion approximation results in percent-level energy scale biases in \Eb with respect to deposited energy above 1.5~MeV, and 10\%-level biases below 1~MeV.  
Blip energy resolution, characterized with MC simulations of uniformly distributed low-energy electrons, is estimated to be less than 10\% above 1~MeV.  

The algorithm matches truth-level MC information to each reconstructed blip, including the identity (PDG number) of the charged particle that produced a majority of a blip's ionization, the parents of that particle, the producing processes of that particle, and more.

\subsection{Signal Blip Selection and Sorting}
\label{subsec:selection}

For this analysis, we apply minimal signal selection cuts to the full population of reconstructed blips in MicroBooNE.  
Blips formed from hits directly adjacent to cosmic muon tracks, such as those induced by low-energy $\delta$~rays, are identified and excluded from consideration. Blips reconstructed on wires directly adjacent to nonfunctional wires are also rejected to reduce inclusion of track fragments or blips with biased reconstructed physics quantities in the signal dataset.

The combined efficiency of blip reconstruction and signal blip selection was determined using a low-energy electron MC dataset with kinetic energies ranging from 0-12 MeV, similar to that described in Ref.~\cite{MicroBooNE:2023ftv}.  
Efficiency is strongly driven at low energies by chosen settings for the \texttt{WireCell} and \texttt{GausHit} modules used to generate reconstructed hits.  
As in Ref~\cite{MicroBooNE:2023ftv}, this study uses specialized datasets processed using lowered settings in these modules that achieved roughly 50\%  blip reconstruction efficiency at 0.5~MeV.  
The efficiency approaches zero below a detected charge of 6,000 electrons ($e^-$), or an effective energy deposition of 250~keV, and reaches a maximum plateau above 80\% around 35,000 $e^-$ (1.5~MeV deposited energy).  
The upper limit of ~80\% is a consequence of nonfunctional TPC wires limiting the sensitive fiducial volume; when this effect is factored out, the maximal efficiency is $>$95\%.

In contrast to prior MicroBooNE blip studies or analyses in other large MeV-scale particle detectors, we refrain from applying selection cuts based on spatial proximity of blips to one another (\emph{i.e.} single-site versus multi-site event topologies in 0$\nu\beta\beta$ or dark matter experiments~\cite{DM_Review,EXO-200:2012pdt,LZ:2015kxe,Majorana:2019ftu,Dunger:2019dfo}) or performing reconstruction of larger multi-site neutrino LArTPC event topologies (\emph{i.e.} multiple-scattering $\gamma$~rays or neutron capture $\gamma$-cascades~\cite{Castiglioni:2020tsu,Fischer:2019qfr}).  
To perform background subtractions or data-driven validations for some aspects of this study, we will segregate and/or compare signal blip datasets based on their location in the TPC, their proximity to cosmic ray tracks, and their energies.  

\section{Datasets and Monte Carlo Simulations} 
\label{sec:datasets}

To perform our study, we use data collected over a 46-day period in June-July of 2018.
For this dataset, readout of the detector was triggered randomly with a function generator during periods where the BNB was not delivering neutrinos to MicroBooNE (referred to in other MicroBooNE publications as `unbiased beam-external' data).  
This is the same dataset used to measure the steady-state presence of radon daughters in MicroBooNE's purified LAr bulk~\cite{MicroBooNE:2023ftv}.  
As mentioned above, this specially processed dataset achieves lower energy thresholds than other MicroBooNE production datasets due to altered signal processing and hit-finding settings.  
The dataset consists of 653,367 total triggered event readouts, equivalent to a total cumulative live time of roughly 35~minutes.  

Portions of the analysis involving study of radiogenic energy depositions requires production of MC datasets in which monoenergetic $\gamma$~rays were generated uniformly in the volume of a subset of the TPC's G10 struts.  
Using \texttt{LArSoft}, three $\gamma$~rays per strut were generated for each event 
and subsequently propagated through the detector using \texttt{Geant4}.  
This level of $\gamma$-ray activity per event, while not realistic, offered faster MC processing times with negligible spatial overlap of blips from unrelated physics processes.  
Simulated LArTPC detector response attributes match those of other MicroBooNE physics analyses~\cite{ub_syst}, with the exception of ionization charge diffusion, which was modeled using diffusion parameters reported in Ref.~\cite{ub_diffusion:}; for further discussion of LArTPC response modeling, see Sec.~\ref{subsec:calib}.  
To ensure the inclusion of realistic wire noise features in addition to simulated MeV-scale energy depositions, MicroBooNE-standard “event overlay” procedures were used, in which the generated wire signatures of the simulated photons mentioned above are added to waveforms from existing unbiased beam-external data events in order to incorporate data-realistic noise in the wire signals. To guarantee only simulated $\gamma$-produced blips were studied in MC datasets, reconstructed blips' truth-level variables were used to discard all real blips present in overlaid events.  
A total of 109,268 overlaid monoenergetic $\gamma$-ray event displays are generated using these methods.  

Studies of cosmogenically-produced MeV-scale energy depositions were aided by production of MC datasets containing simulated cosmic rays.  
The use of the CORSIKA cosmic ray generator to simulate cosmogenic activity in MicroBooNE is discussed in detail in Ref.~\cite{MicroBooNE:2020fmc}.  
For our study, we sample from the contents of CORSIKA showers generated for this previous study. Since that study showed consistency between data and simulation for observed muon rates when CORSIKA simulated only proton shower generation in the upper atmosphere, we maintain this CORSIKA setting for our study. The shower content is generated on a surface 18~m above the cryostat that extends 10~m beyond the LArTPC cryostat in the $x$ and $z$ directions. Initial and secondary particles are propagated from this surface through the detector by \texttt{Geant4}. Cosmic particles are generated randomly in time $\pm $ 2.8 ms around the readout trigger  to account for ionization produced prior to the trigger that can still be collected by the drift readout. As with monoenergetic $\gamma$-ray MC samples described above, CORSIKA MC samples are also overlaid with beam-off external data, and truth-level variables are used to discard consideration of real blips and tracks present in overlaid events.

\begin{figure}[hptb!]    
\includegraphics[trim=0.25cm 0.0cm 0.0cm 0.0cm, clip=true, width=0.48\textwidth]{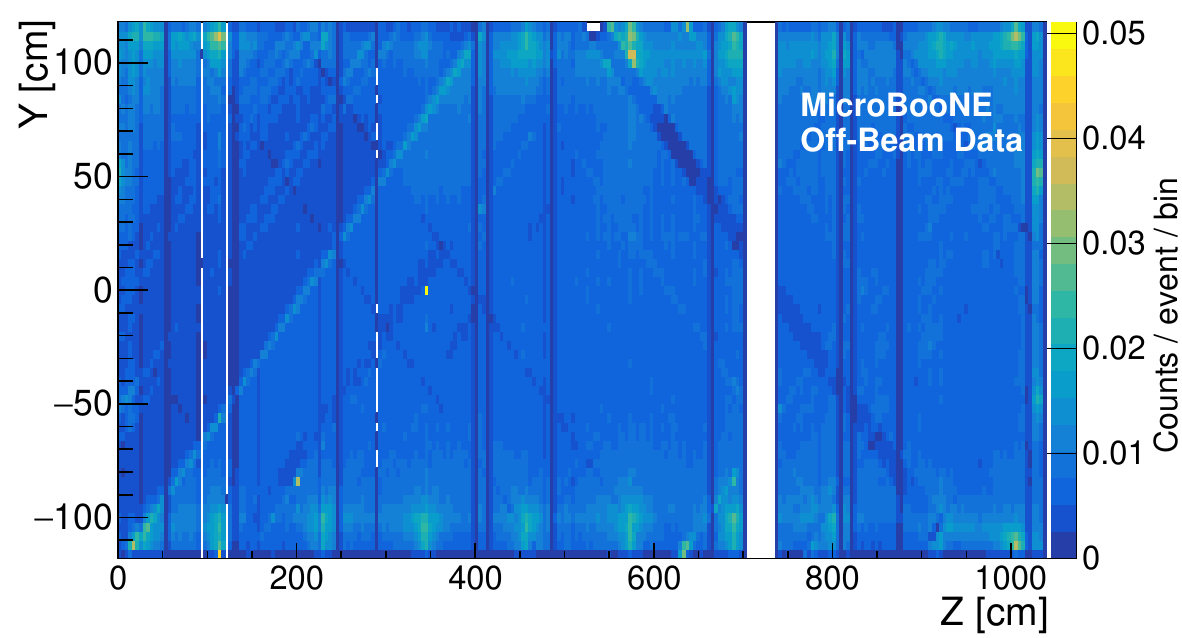} 
\caption{Distribution of reconstructed blip positions per event display in the TPC volume projected onto the $y$-$z$ plane.   
As described in the text, $y$ and $z$ coordinates correspond to the vertical and neutrino beam directions, respectively.}
\label{fig:yz}
\end{figure}

\section{General Description of MeV-Scale Signals}  \label{sec:analysis}

A total of 69,427,354 candidate signal blips were reconstructed, corresponding to an average of 106 blips per event, or roughly 0.39 blips per kg of LAr inside the TPC per second of active detector readout.  
The position distribution of signal blips within the TPC in the $y$-$z$ plane, shown in Fig.~\ref{fig:yz}, exhibits a few visible artifacts due to the presence of non-functional or noisy wires, resulting in diagonal and vertical regions containing more or less blip activity than usual. Distinct regions of elevated blip activity, termed “hot spots”, are visible along the edges of the detector. These closely correspond to the locations of G10 field cage support struts described in Sec.~\ref{sec:microboone} and pictured in Fig.~\ref{fig:TPCdrawing}. This observation supports the hypothesis that the fiberglass laminate composing the strut contains radioactive impurities.
\begin{figure}
\includegraphics[width=0.5\textwidth]{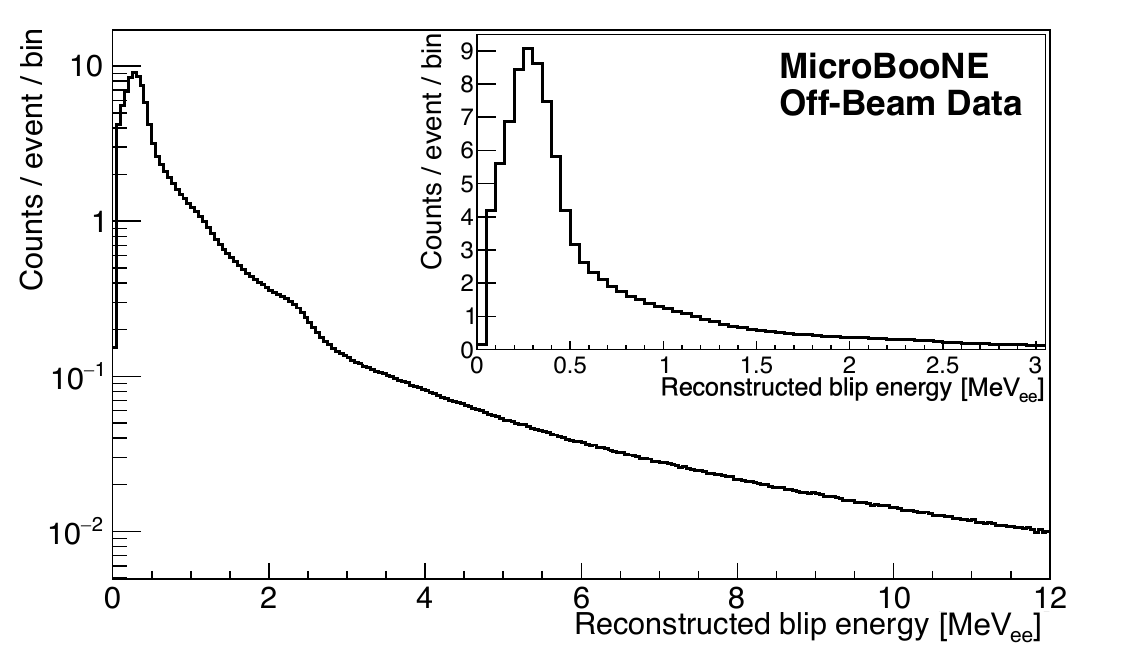}
\caption{Spectrum of reconstructed energies, \Eb, for all signal blips in MicroBooNE cosmic data. The inset shows the same spectrum from 0-3~\MeVee plotted with a linear scale.}
\label{fig:Eblip_spectrum}
\end{figure}
Further insight into ambient blip activity and the observed hot spots is provided by their reconstructed energy (\Eb) spectrum given in Fig.~\ref{fig:Eblip_spectrum}.  
Highest counts appear at the lowest accessible \Eb in MicroBooNE, with a kink in the spectrum occurring near the 0.58~MeV $\beta$-decay endpoint of $^{39}$Ar, a naturally-occurring isotope present in LAr at ($\sim$1~Bq/kg)~\cite{DEAP:2023wri,WARP:2006nsa}.  
The region of the \Eb spectrum below 0.58~\MeVee, containing on average 65 blips per event (roughly 0.24 blips per kg of LAr inside the TPC per second of active detector readout), is heavily sculpted by changes in blip reconstruction efficiency described in Sec.~\ref{sec:datasets}, making it difficult to perform other studies in the absence of precise threshold characterization, modeling, and calibration.   
For this reason, we leave the lower-energy blip population as a topic for future study. 

\begin{figure}

\includegraphics[trim=0.0cm 0.0cm 0.0cm 0.0cm, clip=true, width=0.47\textwidth]{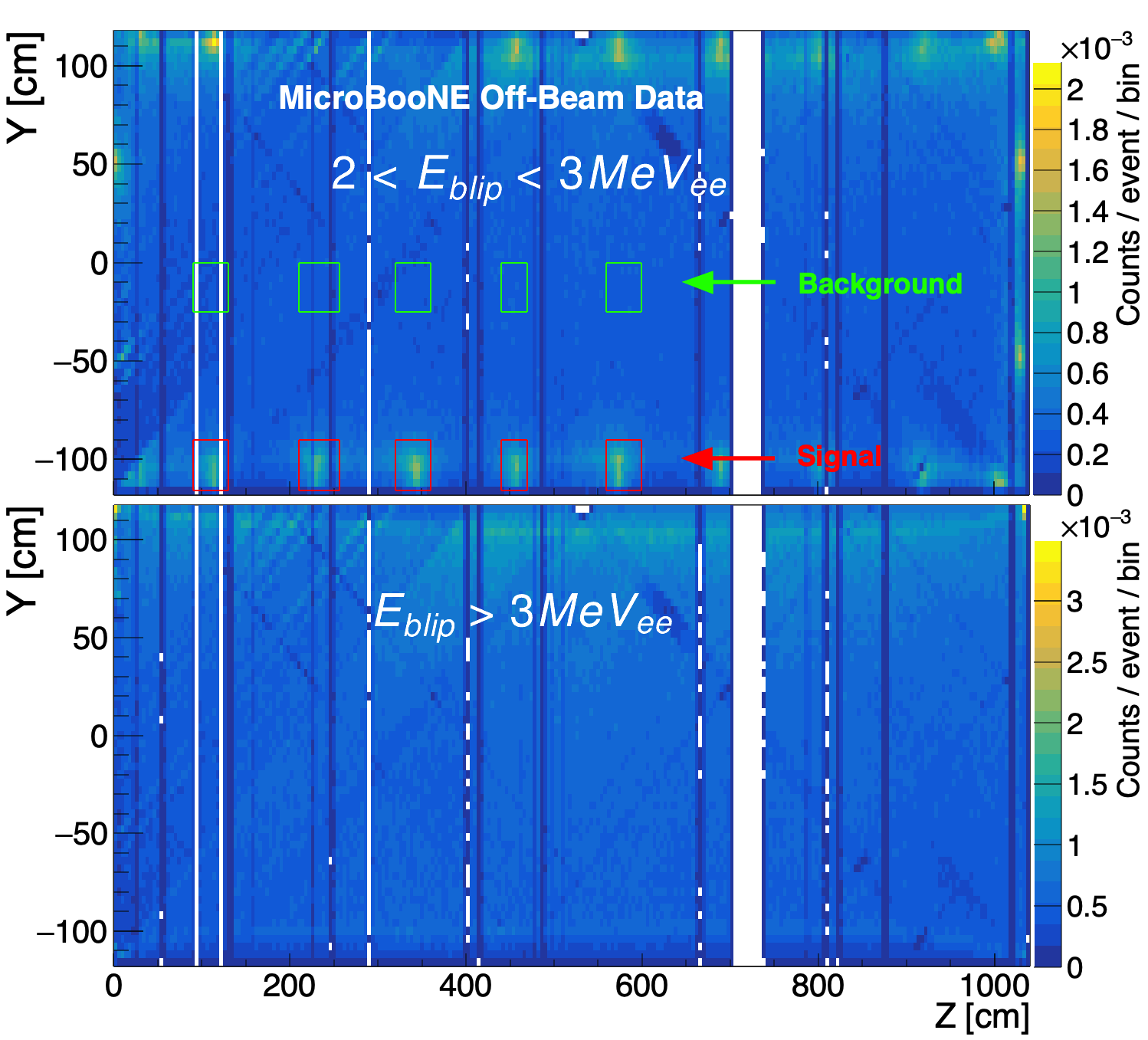}
\caption{Distribution of reconstructed $y$-$z$ positions for blips consisting of matched hit clusters on all three planes, with \Eb between 2 and 3~\MeVee (top) and above 3~\MeVee (bottom).  Red and green boxes in the top panel indicate signal and background regions for blip hot spot studies described in Sec.~\ref{sec:hotspot}.}
\label{fig:yz_2}
\end{figure}

Another obvious energy spectrum feature is present between 2 and 3 \MeVee, a regime well above the blip reconstruction threshold.  
Roughly 4.8 signal blips per event fall into this energy range.  
This spectral feature is generally consistent with a prominent monoenergetic 2.614~MeV $\gamma$-ray associated with $\beta$-decay of $^{208}$Tl, a daughter of the long-lived $^{232}$Th radioisotope~\cite{nndc_chart}.  
Figure~\ref{fig:yz_2} shows the $y$-$z$ distribution of blips occurring in this 2-3 \MeVee range.  
Hot spot regions are still present in the narrower energy range dominated by this spectral feature.  
This suggests that the G10 struts are a source of $^{208}$Tl $\gamma$~rays, and that $^{208}$Tl and its ancestor $^{232}$Th may be present in this material.  
This strong spectral feature in MicroBooNE's blip energy spectrum, combined with its apparent localization in specific MicroBooNE components, offers unique possibilities for LArTPC response calibration that will be studied in detail in the following section.  

Above 3 \MeVee in Fig.~\ref{fig:Eblip_spectrum}, a featureless and exponentially falling energy spectrum is observed, with around 7.8 blips per event in this energy range (4.4 blips with \Eb $>$ 5 \MeVee).  
As shown in the bottom panel of Fig.~\ref{fig:yz_2}, these higher energy blips are not clustered around the G10 struts, but instead are spread evenly throughout the detector, with a greater concentration near the top.  
This position distribution suggests a cosmogenic origin for these blips, which will be explored further in Sec.~\ref{sec:cosmic}.

\begin{figure}

\includegraphics[trim=0.0cm 0.0cm 0.0cm 0.0cm, clip=true, width=0.47\textwidth]{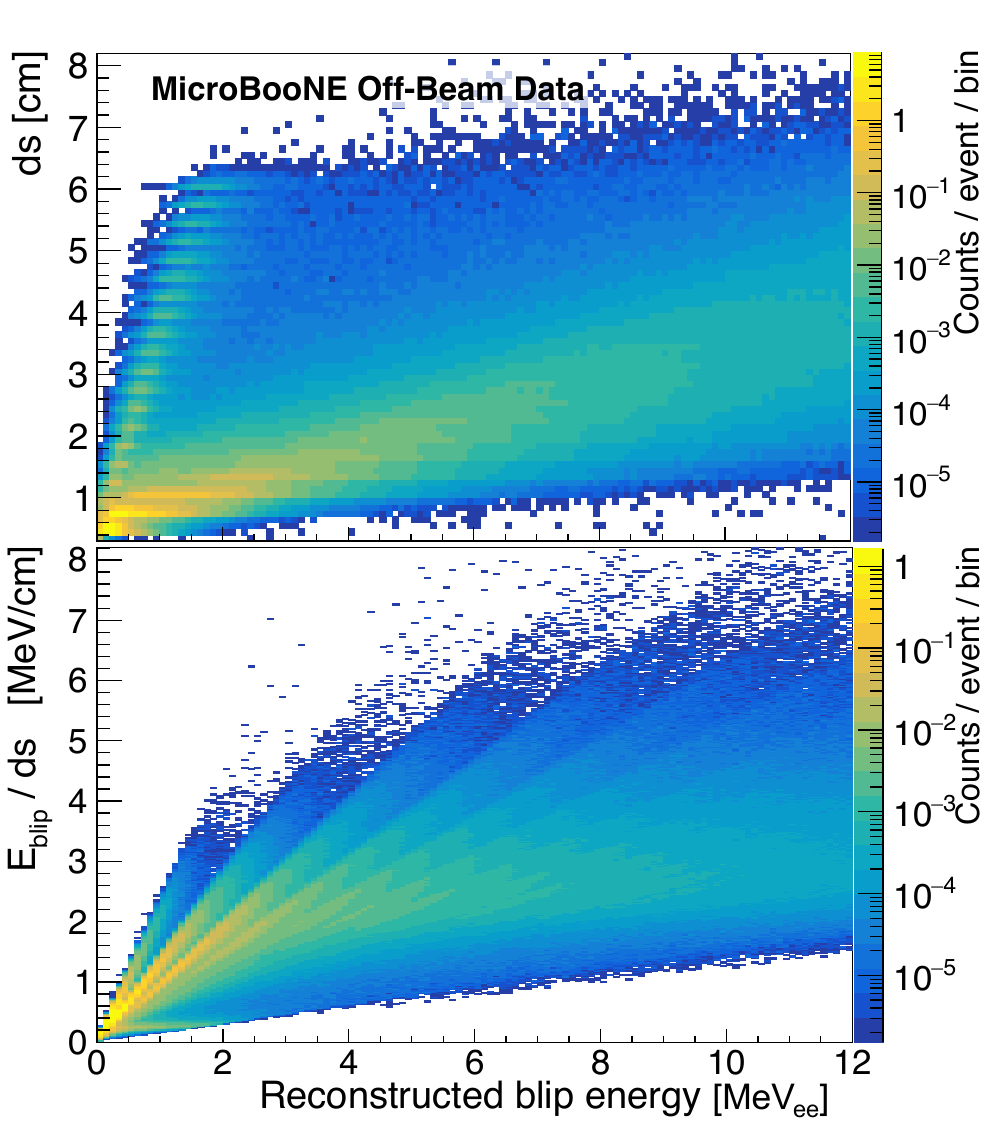} 
\caption{Distributions of reconstructed blip size $ds$ (top), and average blip energy per unit length $E/ds$ (bottom), plotted against the reconstructed blip energy. Striped patterns present in these distributions are artifacts of the discrete spacing of MicroBooNE's charge collection elements. The smaller secondary population extending to high~$ds$ in the top panel corresponds to blips featuring incorrectly plane-matched clusters.}
\label{fig:blip_size_2}
\end{figure}

Reconstructed blip size is visualized in Fig.~\ref{fig:blip_size_2}, showing a clear positive correlation between a blip’s estimated length $ds$ and its energy.  
The secondary population at high $ds$ and low \Eb~are blips featuring plane-matched clusters generated by different physics processes (or `bad plane matches'), which represent roughly 1\% of the total reconstructed blip population.  
This figure also depicts blips’ $E/ds$, which is analogous to the stopping power ($dE/dx$) typically reconstructed for extended tracks in LArTPCs. Artifacts due to the use of discretized charge readout elements are present in both distributions in Fig.~\ref{fig:blip_size_2}, an unavoidable consequence of studying small energy depositions with physical lengths on the order of the wire-to-wire separation.  The substantial spread is also caused by the meandering, stochastic nature of low-mass charged particle trajectories in matter (well illustrated in, for example, Ref.~\cite{NEXT:2018zho}) and of ionization charge-sharing between wires on a plane.  

In Sec.~\ref{sec:cosmic}, we will study further whether particles with differing average $dE/dx$, such as protons and electrons, can be distinguished using blip size variables.  

\section{Study of Radiogenic Hot Spots}
\label{sec:hotspot}

The 2-3 \MeVee spectral feature present in the vicinity of MicroBooNE's G10 struts calls for further study.  
In this section, we use this feature to perform the first percent-level-precision energy scale calibration of a neutrino LArTPC in the MeV regime and to measure the specific activity of $^{208}$Tl in MicroBooNE's G10 struts.  Prior to carrying out these studies, our hypothesis regarding the physics origin of this spectral feature was further verified using the single-$\gamma$ MC datasets described in Sec.~\ref{sec:datasets}. 

\begin{figure}[tb]
\includegraphics[trim=0.0cm 0.0cm 0.0cm 0.5cm,, width=0.5\textwidth]{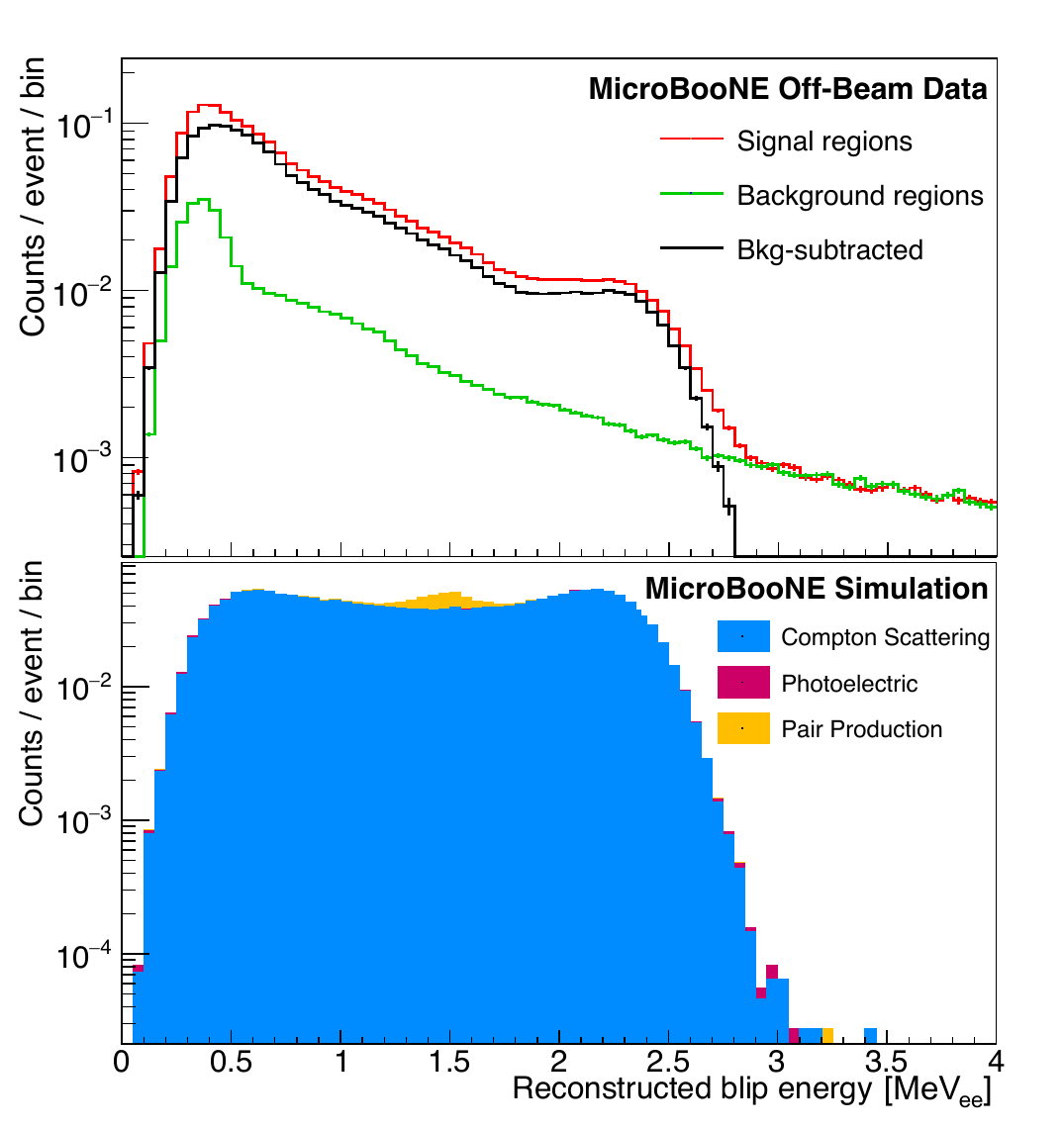}
\caption{\Eb spectra (top) for data blips located in the signal and background regions defined in Fig.~\ref{fig:yz_2}, as well as their difference. Background-subtracted \Eb spectrum (bottom) in the same TPC regions produced by simulated monoenergetic 2.614~MeV $\gamma$~rays in nearby G10 struts with stacked components by process.}
\label{fig:Tl208}
\end{figure}

Potential issues related to dead wires were reduced by examining and simulating $\gamma$-ray emissions from a subset of five struts on the detector bottom, highlighted in Fig.~\ref{fig:yz_2}.  
To study characteristics of only blips related to the G10 struts, we exclude all signal blips within 15~cm of a track, and then implement a background-subtraction procedure, with reconstructed signal and background blip locations indicated by the red and green boxes in Fig.~\ref{fig:yz_2}.  
For both data and simulation, the energy spectrum of the green boxed sample far from the strut is then subtracted from that of the red boxed sample adjacent to the strut.  
Figure~\ref{fig:Tl208} depicts the reconstructed energy spectrum for data blips in the signal and background boxes as well as the subtracted result.  
The result of the subtraction is an increase in the prominence of the edge feature between 2 and 3~\MeVee~\Eb, mostly due to the reduction in content above roughly 2.5~\MeVee.

The background-subtracted blip energy spectrum from the $\gamma$-ray MC dataset is also depicted in Fig.~\ref{fig:Tl208}.  
In the 2-3 \MeVee range, the simulated spectrum contains a very similar feature to the data spectrum, \textit{i.e.} the falling edge.
This feature in MC is due almost entirely to Compton-scattered electrons produced by the simulated 2.614~MeV $^{208}$Tl decay $\gamma$ rays from the G10 struts. Thus, the feature of interest found in the data's \Eb spectrum is likely a Compton edge produced by interaction of 2.614 MeV $\gamma$~rays generated by $^{208}$Tl radioactive decays in MicroBooNE's G10 struts.  

It should be noted in Fig.~\ref{fig:Tl208_BestFit}, that the background-subtracted \Eb spectrum contains substantial content below the edge feature, and also that data and MC spectra diverge widely below roughly 2~MeV.  
This is likely due to the presence of other radioisotopes in the decay chains of thorium and uranium that are also present in the struts.  
A closer analysis of this lower-energy portion of the \Eb spectrum requires accurate simulation of these isotopes and their various decay branches, a task that is beyond the scope of the current study.  
Fortunately, as demonstrated in a wide range of large  MeV-scale particle detectors~\cite{DayaBay:2019fje,Borexino:2009mcw,EXO-200:2014ofj,PROSPECT:2020sxr}, the 2.614~MeV $^{208}$Tl $\gamma$-ray's uniquely high energy and high intensity (produced in $>$99\% of $^{208}$Tl decays) allows its Compton edge and full-energy peak to appear in measured ambient background spectra with little interference from other radioisotopes.  

\subsubsection{MeV-Scale Energy Calibration Demonstration}
\label{subsec:calib}

\begin{figure}
\includegraphics[width=0.49\textwidth]{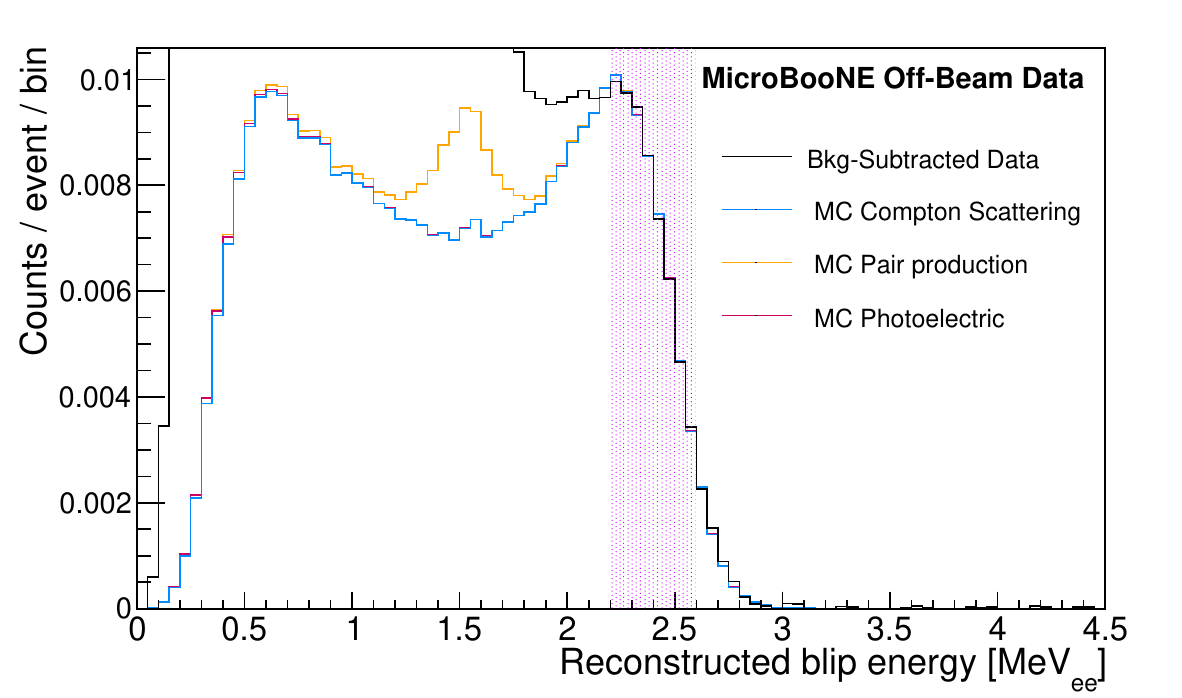}
\caption{Comparison of the measured Compton edge from 2.6~MeV  $^{208}$Tl decay $\gamma$~rays to MC simulations of this feature with best-fit energy scale shift and specific activity parameters applied. Fit range indicated by the pink shaded region, see the text for further details about fit parameters and approach.}
\label{fig:Tl208_BestFit}
\end{figure}

With simulated and measured $^{208}$Tl $\gamma$-ray Compton edges now clearly defined, we perform a calibration of reconstructed blip energy scales by testing the alignment of this feature between data and MC.  
Alignment is achieved by simultaneously applying an energy scale correction and a normalization scaling to the simulated spectrum, with best alignment judged by calculating the $\chi^2$ between the data and adjusted simulated spectrum between 2.2 and 2.6~\MeVee.  
As shown in Fig.~\ref{fig:Tl208_BestFit}, we scale the energy of simulated events up by 3.12\% in order to best match the data ($\chi^2$/ndf of 2.02/6), ndf being the number of bins minus the number of parameters used for the fit. 
Marginalizing over the normalization parameter, a $\Delta \chi^2$ below 2 was observed for an energy scaling parameter between 2.47\% and 3.57\%, defining the 1$\sigma$ statistical uncertainty for that parameter.  

A variety of systematic uncertainties, listed in Table~\ref{tab:uncertainties}, have the potential to bias the simulated \Eb spectrum relative to data. The impacts of most parameter uncertainties are benchmarked by performing simulations with that parameter adjusted to the edge of its 1$\sigma$ allowed region, refitting the data, and using the new best-fit value to define its 1$\sigma$ systematic uncertainty contribution.  
Some parameters, such those related to diffusion, recombination, and space charge detector response parameters, have been well-established by previous MicroBooNE analyses~\cite{ub_syst,ub_diffusion:}.  
For diffusion, the longitudinal diffusion parameter $D_L$ and its uncertainty were taken from Ref.~\cite{ub_diffusion:}, while the transverse diffusion parameter $D_T$, as done in Ref.~\cite{MicroBooNE:2023ftv}, was assigned a large 30\% uncertainty given the lack of existing measurements; $D_L$ and $D_T$ uncertainties were assumed to be fully correlated.  
Recombination uncertainty was assessed by comparing results between the Birks and modified box models of ionization quenching~\cite{argo_modbox,ub_cal}.  
For space charge, we use maps corresponding to 1$\sigma$ uncertainty ranges as defined in Refs.~\cite{Abratenko_2020,ub_syst}. Due to the high LAr purity in MicroBooNE during the period used for this analysis, ionization electron drift losses and associated uncertainties were negligible. Other uncertainties, such as that of the size and placement of struts, are conservatively estimated using mechanical drawings and photographs from the period of MicroBooNE detector construction.  
Background subtraction uncertainties were conservatively benchmarked using the case where no background subtraction is applied.  
Finally, minimization is performed with modest adjustments to the fit ranges to benchmark uncertainties related to the specific fit approach.  
The quadrature-summed total systematic uncertainty from all sources is also given in Table~\ref{tab:uncertainties}. The approach followed to evaluate uncertainties is tailored to this MeV-scale analysis, and differs in several points to the data-driven detector systematic estimation detailed in~\cite{ub_syst} and used in MicroBooNE’s higher energy measurements.

\begin{table*}[tb!]
\begin{tabular}{|l|c|c|c|}
    \hline
    \multirow{3}{*}{Parameter} & \multirow{3}{*}{Value and Uncertainty} & \multicolumn{2}{c|}{Deviation from CV}  \\
    & & Energy Scale  & G10 Activity \\
    & & (absolute, \%)    & (relative, \%) \\
    \hline
    
    Strut mass & Nominal strut dimensions versus $\pm$1~mm  & $<$0.2 & 7.1 \\ 
    Strut extent into active volume & ($1.2 \pm 1.0$)~cm & $<$0.2 & 18.1 \\ 
    Recombination & Modified box versus Birks & $<$0.2 & $<$2.0 \\ 
    Space charge & Default versus $\pm$1$\sigma$ charge map & 0.3 & $<$2.0 \\ 
    Diffusion & $D_{L}= 3.74 ^{+0.28}_{-0.29} , D_{T}=5.9^{+1.8}_{-1.8}$ (cm$^2$/s) & $<$0.2 & $<$2.0 \\
    Bin width  & 0.05~MeV versus 0.1~MeV & 0.02 & 0.1 \\
    Fit range & $2.2^{+0.3}_{-0.1} - 2.6^{+0.2}_{-0.4}$ MeV & 0.6 & 12.4 \\ 
    Background subtraction & Default versus no subtraction  & 0.91 & 13.5 \\ 
    \hline
    \textbf{Total} &  & \textbf{1.2} & \textbf{26.9} \\ \hline

    \end{tabular}
\caption{Summary of systematic uncertainties considered in \Eb energy scale calibration and $^{208}$Tl specific activity measurements.  For each uncertainty source, the central value (CV) and bounding scenario cases are given, along with the absolute change in the energy scale correction and fractional change in activity when the parameter is varied in the specified manner.  Statistical uncertainties associated with the MC-derived energy scale and activity changes are 0.2\% and 2.0\%, respectively.
The extent of the G10 strut protruding past the field cage profiles into the active LAr volume is visualized in Fig.~\ref{fig:TPCdrawing2}.  See the text for more description of each systematic error source and its applied adjustment.}
\label{tab:uncertainties}
\end{table*}

Thus we find that the MC energy scale correction parameter providing the best fit between MicroBooNE data and MC simulations is (3.12~$\pm$~0.2~\text{(stat)}~$\pm$~1.2~\text{(syst)})\%. This MeV-scale calibration using ambient radiological activity is the first of its kind for a large neutrino LArTPC. 
The general level of correspondence between data and simulation indicates that MeV energy scales in MicroBooNE are well-modeled to the few-percent level.  
This result improves upon previous demonstrations in MicroBooNE performed using $^{214}$Bi $\beta$-decays~\cite{MicroBooNE:2023ftv} by improving blip-based energy calibration precision to the percent level.

A blip-based calibration scheme similar to the one demonstrated here, potentially using the same $^{208}$Tl Compton edge feature, could serve as an attractive option for detector-wide, region-specific, or wire-by-wire energy scale calibrations in future large LArTPCs, particularly those deployed underground in low cosmic ray environments, such as DUNE.  
The non-uniform distribution of $^{208}$Tl inside the detector may limit its attractiveness as a calibration source: for example, the present analysis has provided a calibration of five detector regions comprising a small fraction of MicroBooNE's active TPC volume.  
In addition, the lack of knowledge of blip true $x$ locations represents a current limitation of blip-based calibration methods more generally.  

\subsubsection{Measurement of G10 Specific Activity}
\label{subsec:G10}

The same fit procedure used to evaluate energy scale agreement between data and MC in the last section can also be used to measure the specific activity of $^{208}$Tl in MicroBooNE's G10 struts.

In the MC, three isotropically-oriented $\gamma$~rays originating at random points within each G10 strut's volume were simulated during each 3.2~ms detector readout, equivalent to a G10 specific activity of 60.8~Bq/kg. 
As shown in Fig.~\ref{fig:Tl208_BestFit}, a best fit between measured and simulated Compton edges occurs when the level of simulated $\gamma$-induced blip content per event is scaled by a factor of 0.193, to 11.7~Bq/kg.
An examination of nearby fit phase space similar to that described in the previous sub-section gives a 1$\sigma$ statistical uncertainty range of $\pm$0.18~Bq/kg, and an identical systematic error determination is also performed, with results given in Table~\ref{tab:uncertainties}.   
The resulting measured specific activity of the G10 struts is $(11.7 \pm 0.2~\text{(stat)} \pm 3.1~\text{(syst)})$~Bq/kg, with the systematic error dominated by uncertainties in the strut’s exact location. This activity level is comparable to existing measurements of $^{208}$Tl and $^{232}$Th activity in fiberglass detector components collected in the community-standard radiopurity database~\cite{radiopurity}, specifically fiberglass component entries from Refs.~\cite{Lawson:2011zz,BOREXINO:2001bob}.  
This database also includes measurements of $^{208}$Tl and $^{232}$Th activity well below that measured for MicroBooNE (for example, other fiberglass components in Ref.~\cite{Lawson:2011zz}), indicating a large potential variability in radiopurity.  
Thus, care should be taken in the selection of fiberglass components, such as structural beams, circuit boards, and other electrically insulating or resistive components, for future LArTPC experiments aiming to perform MeV-scale physics measurements.  
Radiopurity screening measurement campaigns may be advisable.  

\section{Study of Cosmogenic Blips}
\label{sec:cosmic}

\begin{figure*}
\includegraphics[trim=0.0cm 0.0cm 1.7cm 1.7 cm, clip=true, width=0.47\textwidth]{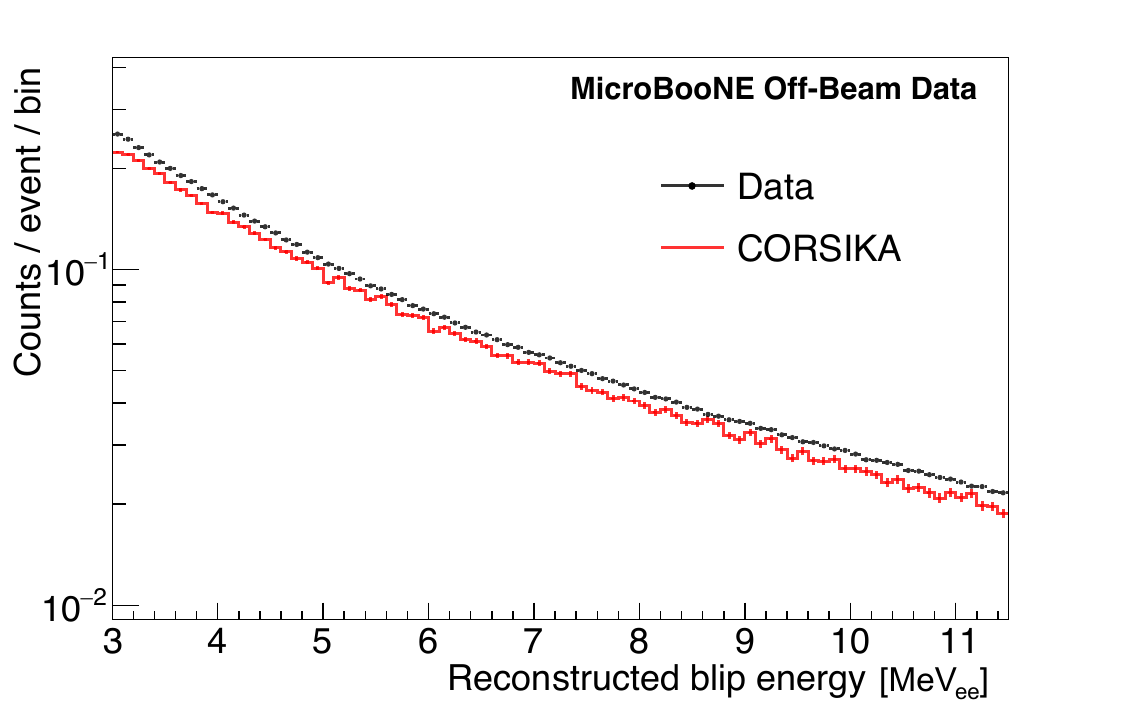}
\includegraphics[trim=0.0cm 0.0cm 0.0cm 0.0cm, clip=true,width=0.51\textwidth]{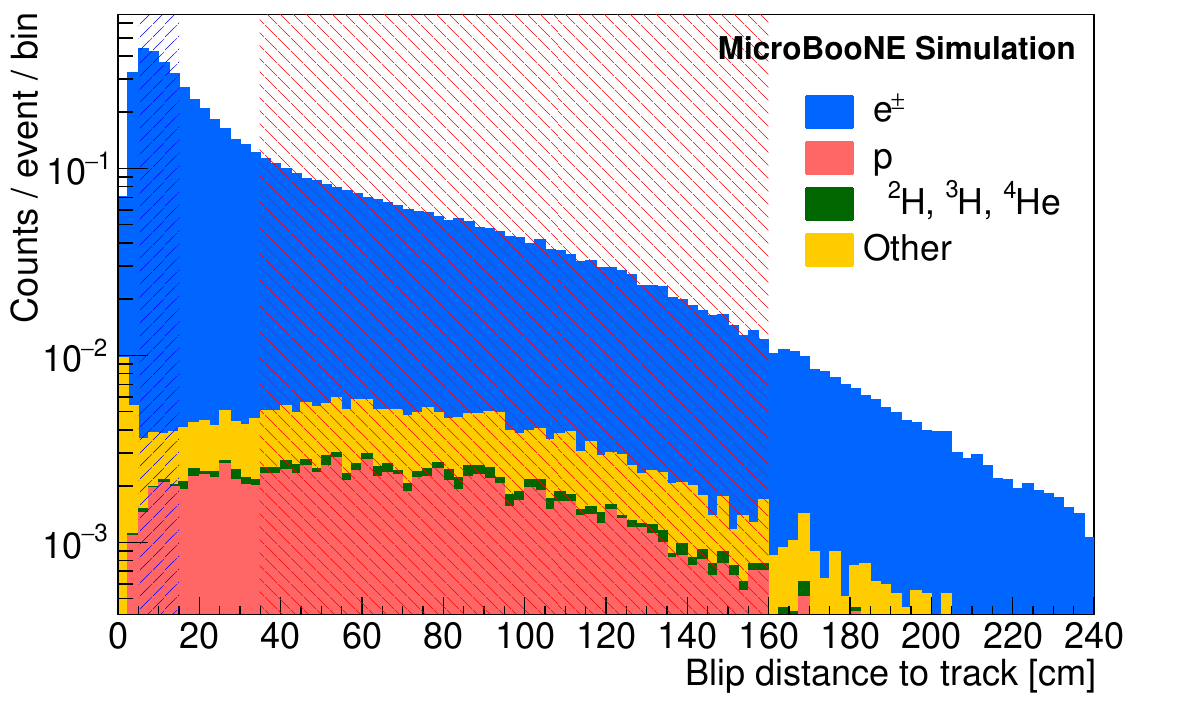}
\caption{(Left) reconstructed \Eb spectrum above 3\,MeV for data (black points) and for CORSIKA MC (red). Error bars indicate statistical uncertainties. (Right) Blip-track proximity, or the distance of closest approach between blips with \Eb$>$3\,MeV and the nearest track longer than 5~cm for CORSIKA MC.   Contributions from different blip parent particle ID classes are shown as stacked colored histograms. Besides $p$, and $e^{\pm}$, heavier nuclei also contribute to reconstructed blips. The \textit{Other} category is comprised primarily by muons/pions indicating blips due to misreconstruction of tracks. Blue (red) transparent vertical bands indicate ranges selected for the short (long) blip-track proximity subsets described in the text.  }
\label{fig:cosmicEDist}
\end{figure*}

We now consider the blip population in Fig.~\ref{fig:Eblip_spectrum} with \Eb$>$ 3 \MeVee, which are distributed throughout the active volume but more concentrated at the top, suggesting a cosmogenic origin.
As shown in Fig.~\ref{fig:cosmicEDist}, the energy spectrum shape in this range is similar between data and the CORSIKA cosmic ray MC described in Sec.~\ref{sec:datasets}.  
While the shape is similar, we observe a 10\% offset in overall normalization between the two, contrasting with the few-percent-level agreement in measured and modeled cosmic muon track rates we previously reported~\cite{MicroBooNE:2020fmc}.  
We observe the same data-MC correspondence in muon rates when considering only long ($>$5cm) tracks, but we see a substantial data-MC offset for short ($<$5cm) tracks that mirrors the aforementioned offset in overall data and MC blip activity.  
We reserve further discussion of observed data-MC mismatches in the cosmic ray flux for Sec.~\ref{subsec:protons}.

Figure~\ref{fig:cosmicEDist} also shows a distribution of track proximity for all blips, defined as the 3D distance of closest approach between each blip and the reconstructed cosmic track $>$5~cm in length nearest to it.   
The shape of the blip-track proximity distribution provides clues to the physics processes generating the cosmic blip population.  
The strong peak at short proximity and subsequent exponential decrease within 10 to 35~cm suggests that a  majority of selected cosmic blips are the byproduct of bremmstrahlung $\gamma$ rays generated following hard scattering of cosmic muons or production of high-energy $\delta$ electrons. This hypothesis is supported by probing the true parentage of CORSIKA MC blips at track proximities between 5 and 15~cm, which is also pictured in Fig.~\ref{fig:cosmicEDist}: an overwhelming majority of these blips ($\gg$99\%) have electrons (from $\gamma$-ray Compton scattering or pair-production) or positrons (from pair production) as their direct producer, and $>$85\% of these true electron parents descend from bremsstrahlung photons generated by $\mu^{\pm}$ $\delta$~rays.    

According to Fig.~\ref{fig:cosmicEDist}, blips farther from tracks (long blip-track proximity) are also primarily produced by true electrons and positrons.  
Within the blip-track proximity window of 35 to 160~cm, 93\% of blips have an $e^\pm$ direct producer.  
Roughly 40\% of these share the same $\delta$-ray bremsstrahlung ancestry that dominated the short (5-15~cm) blip-track proximity sample, indicating that a substantial fraction of the diffuse, track-uncorrelated blips in MicroBooNE are produced by muons whose trajectories are largely or entirely outside the active TPC volume. Blips from $\gamma$ rays from external cosmic muons appear randomly in time within the TPC readout period, so their measured $x$ coordinate cannot be used for fiducial-based mitigation along the drift direction.  
The other primary parentage categories are cosmic neutron inelastic nuclear scattering (roughly 30\%) and electromagnetic showers not associated with muon activity (roughly 25\%).  
Absent further information from other detector systems, such as a muon tagger or an efficient light collection system, or cuts on other LArTPC variables, such as nearby shower objects or high adjacent blip activity, these blip categories represent potentially troublesome backgrounds to purely MeV-scale topologies for future low-energy neutrino measurements in near-surface LArTPC experiments~\cite{Grant:2015jva, Gardiner:2021qfr, Asaadi:2022ojm, Aguilar-Arevalo2023dai}.   

Figure~\ref{fig:cosmicEDist} also shows a sub-dominant but substantial population of blips in CORSIKA simulations that is directly produced by protons.  
While representing only 0.3\% of the population at short (5-15~cm) blip-track proximity, proton-generated blips make up 5\% of the predicted population at long blip-track proximity (35-160~cm).  
A vast majority of blip-generating protons, $>$99\%, have neutrons as their direct producer.  
This indicates that this population is generated by inelastic $(n,p)$ interactions~\cite{PhysRevC.86.041602} between incident cosmic neutrons and argon nuclei in the active TPC.  

\subsubsection{Defining Particle Discrimination Criteria}
\label{subsec:pid}

\begin{figure}
\includegraphics[width=0.48\textwidth]{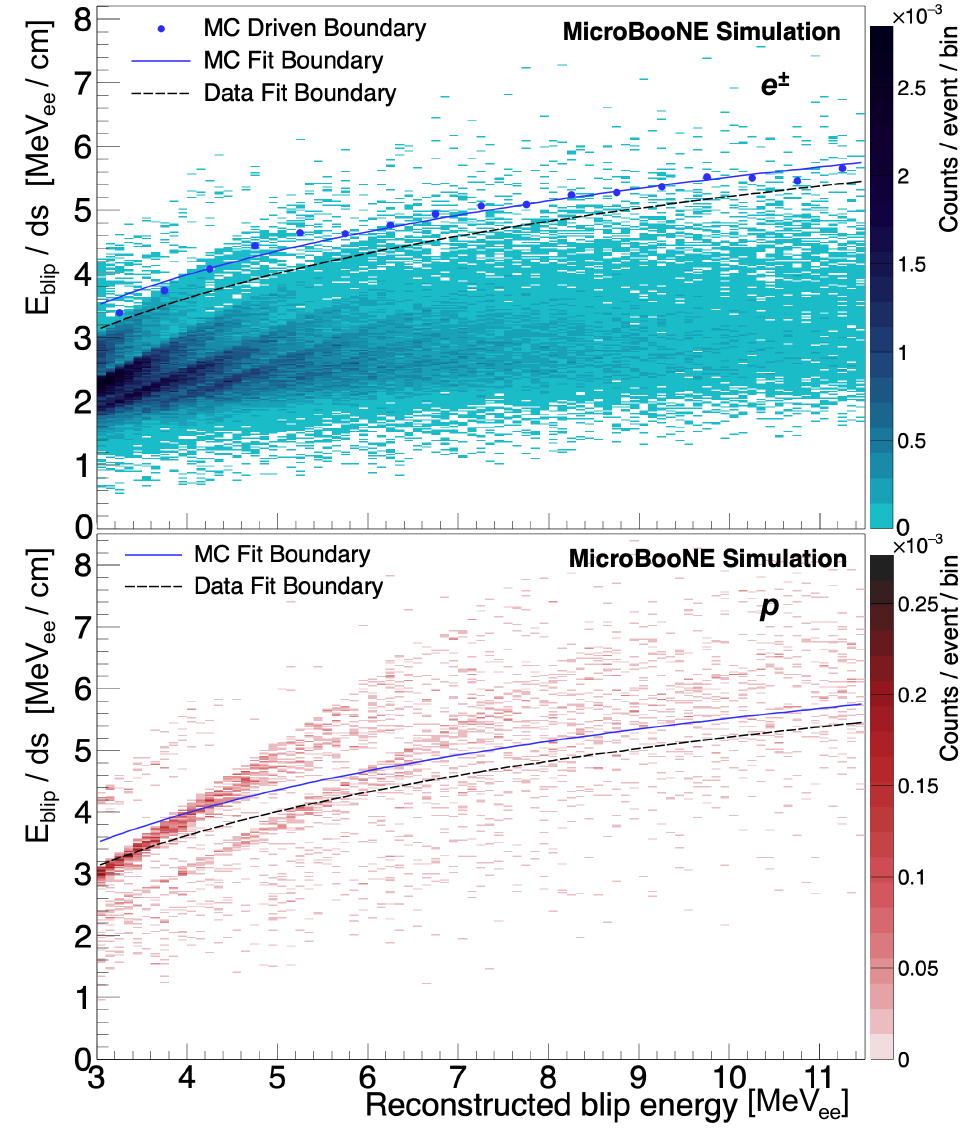}
\caption{Distribution of average energy deposition density, \Eb/$ds$, for blips of various reconstructed energies in CORSIKA MC simulations. Blips generated by true $e^\pm$ parents are shown in blue (top), while those with $p$ parents are shown in red (bottom). In the top panel, the blue points show the \Eb/$ds$ value above which 1\% of CORSIKA MC blips reside for each $0.5~\text{MeV}_\text{ee}$  bin. The blue and black lines show the best-fit log function (Eq.~\ref{eq:PID}) for MC and data, respectively.}
\label{fig:PID}
\end{figure}

Figure~\ref{fig:PID} shows reconstructed average energy deposition density, $E_\text{blip}/ds$, versus \Eb for the long blip-track proximity (35-160~cm) sample in CORSIKA MC, separated into two direct producer categories: $p$ versus $e^\pm$. 
The distributions, while overlapping, are clearly offset from one another, indicating that MicroBooNE should be capable of exploiting the  offset in stopping power of low-energy charged particles of differing masses.  
While an offset might also be expected between the two prominent electromagnetically-produced blip categories  -- pair-production (an $e^+e^-$ pair) and Compton scattering (a single $e^-$) -- we find this offset to be fairly small in reconstructed blip space and do not consider this comparison further.  

To demonstrate $p$-$e^\pm$ discrimination, we outline a particle identification (PID) cut based on the value of $E_\text{blip}/ds$ below which 99\% of all blips reside for each $0.5~\text{MeV}_\text{ee}$ bin. Discrete boundary values provided by the comparatively $e^\pm$-pure sample of CORSIKA MC blips with short (5-15~cm) blip-track proximities are illustrated by the blue points in the top panel of Fig.~\ref{fig:PID}.    
A continuous PID selection boundary is then defined by a log fit applied to this set of per-sub-range boundary values: 
\begin{equation}
\label{eq:PID}
    f(E_\text{blip}) = a_1 \ln( a_2 E_\text{blip}).
\end{equation}  
Figure~\ref{fig:PID} also depicts the log function providing the best fit to the short blip-track proximity sample for data and for CORSIKA MC.  For CORSIKA MC, the fit defines a continuous rejection cut boundary at $(a_1, a_2) = (1.69~[\text{MeV}_\text{ee}\text{/cm} ], 2.64~[\text{MeV}_\text{ee}^{-1} ])$.
While 1$\sigma$ statistically allowed ranges for the parameters, $(\pm0.08~[\text{MeV}_\text{ee}\text{/cm} ], \pm0.35~[\text{MeV}_\text{ee}^{-1} ])$, are wide, this is mostly due to fit degeneracy.  If $a_1$ is held constant at its best-fit value, the 1$\sigma$ range for $a_2$ is reduced to $\pm0.04~[\text{MeV}_\text{ee}^{-1} ]$.

\begin{figure}
\includegraphics[trim=0.0cm 0.0cm 0.0cm 0.0cm, clip=true, width=0.5\textwidth]{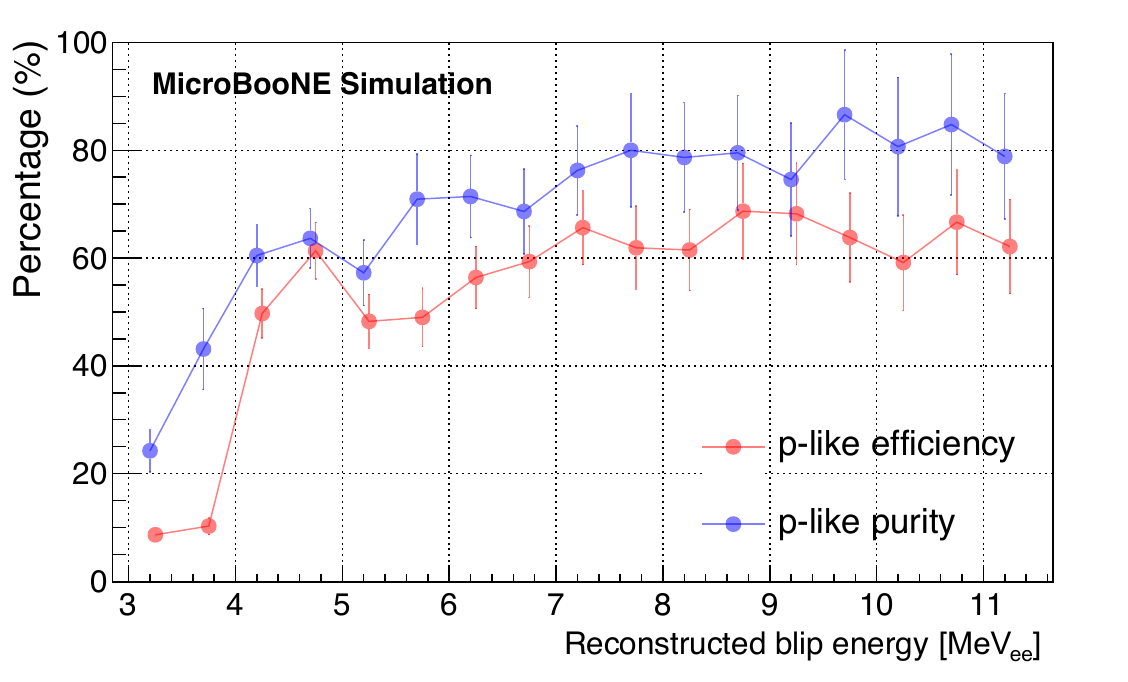}
\caption{Red: Efficiency for selecting reconstructed blips produced by true hadronic parents with the PID cut described in the text.  Blue: composition of the reconstructed blip sample selected with this PID cut.  The hadronic sample groups together contributions from $p$, $^2$H, $^3$H, and $^4$He, with the latter three comprising roughly 7\% of the selected sample.  Error bars represent statistical uncertainties in the CORSIKA MC sample.}
\label{fig:PID_Applied}
\end{figure}

In Fig.~\ref{fig:PID_Applied}, we show the results of applying this PID selection to the long blip-track proximity CORSIKA MC sample shown in Fig.~\ref{fig:PID}.  
The efficiency of the PID cut, defined as the number of reconstructed hadron-generated blips above the cut value over the total number of reconstructed blips, is shown for the $p$-like category as well as its fractional contribution to the total selected sample.  
Blip contributions from other nuclear fragments ($^2$H, $^3$H, and $^4$He) are included in the $p$-like sample, and comprise roughly 7\% of the total.  
The efficiency of the proton-like PID cut starts at roughly 10\% near 3~\MeVee, but quickly rises to 50\% around 4~\MeVee and reaches a maximum of roughly 70\% at higher \Eb.  
We note here that 4~\MeVee \Eb corresponds to roughly 18~MeV true proton energy; this conversion is detailed more fully in subsequent sections, and in the supplementary materials accompanying this manuscript.  

Sample purity, defined as the number of selected true hadron-generated blips divided by the total number of blips above the PID cut value, is also shown in Fig.~\ref{fig:PID_Applied}.  
The applied cut yields a candidate proton blip sample that has $>$50\% purity starting at around $3.75~\text{MeV}_\text{ee}$ \Eb, which increases to $>$80\% at the highest considered blip energies.  
Even at the lowest considered \Eb, where proton selection efficiency is low, the cut nonetheless provides roughly an order of magnitude reduction in electron-like backgrounds relative to the proton-like sample.  
This suggests that meaningful $p/e$ discrimination is possible at and even below 20~MeV in true proton energy.  

When the demonstrated PID cut technique and boundary is applied to MicroBooNE data, statistical and systematic uncertainties related to these quoted efficiencies and background rejection factors must be considered and propagated.  
To propagate statistical uncertainties, we use the fit uncertainty in $a_2$ while fixing $a_1$ to address the fit degeneracy described above.    
A check of this procedure fixing $a_2$ and allowing $a_1$ to float yields similar allowable statistical variation in the PID boundary.
We define systematic uncertainty contributions using similar techniques to that described in Sec.~\ref{sec:hotspot}: the PID cut boundary is rederived and applied using systematically varied MC datasets.    
Considered systematic effects include electron and proton recombination, diffusion, space charge, blip energy scales as benchmarked in Sec~\ref{subsec:calib}, and signal impurity.  
A blip energy scale uncertainty is defined using the difference in results between the CORSIKA dataset with 0\% and the best-fit +3.12\% applied energy scale shift from Sec.~\ref{subsec:calib}, while a signal impurity systematic is defined by the difference in PID cut boundary generated by the full short blip-track proximity dataset versus only the subset of true $e^{\pm}$-produced blips.  
Systematically varied recombination, diffusion, and space charge datasets are defined as described in Table~\ref{tab:uncertainties}.   
Recombination and energy scale systematics provide the dominant uncertainty to particle identification measurements, while MC statistical uncertainties also serve as a non-negligible contributor.  
The impacts of systematic uncertainties on reported proton counts are described in the following subsection.

\subsubsection{Identification of Cosmogenically Produced Protons in MicroBooNE}
\label{subsec:protons}

With the use and performance of the PID cut variable now defined using CORSIKA MC, we apply a data-derived PID cut to the MicroBooNE dataset described in Sec.~\ref{sec:datasets}.  
Using the short blip-track proximity dataset in data, we observe best-fit PID boundary parameters of $(a_1, a_2) = (1.72\pm 0.03~[\text{MeV}_\text{ee}\text{/cm} ], 2.08\pm 0.10~[\text{MeV}_\text{ee}^{-1} ]) $.  When setting $a_1$ to its best-fit value, we find an uncertainty on $a_2$ of $\pm0.01~[\text{MeV}_\text{ee}^{-1}]$.
As shown in Fig.~\ref{fig:PID}, substantial differences in PID boundary fit constants are observed between data and CORSIKA MC.  
These differences are likely related to the mismatch between modeled and measured wire pulse shapes, which are illustrated and discussed in Ref.~\cite{MicroBooNE:2018vro}.  
While this offset is notable, its impact on PID systematic uncertainties is mitigated by using the data's $e^{\pm}$-rich sample at short blip-track proximity to define PID regions for other proton-rich blip subsets in data, as opposed to applying an MC-derived PID boundary directly to the data.  

\begin{figure}
\includegraphics[width=0.5\textwidth]{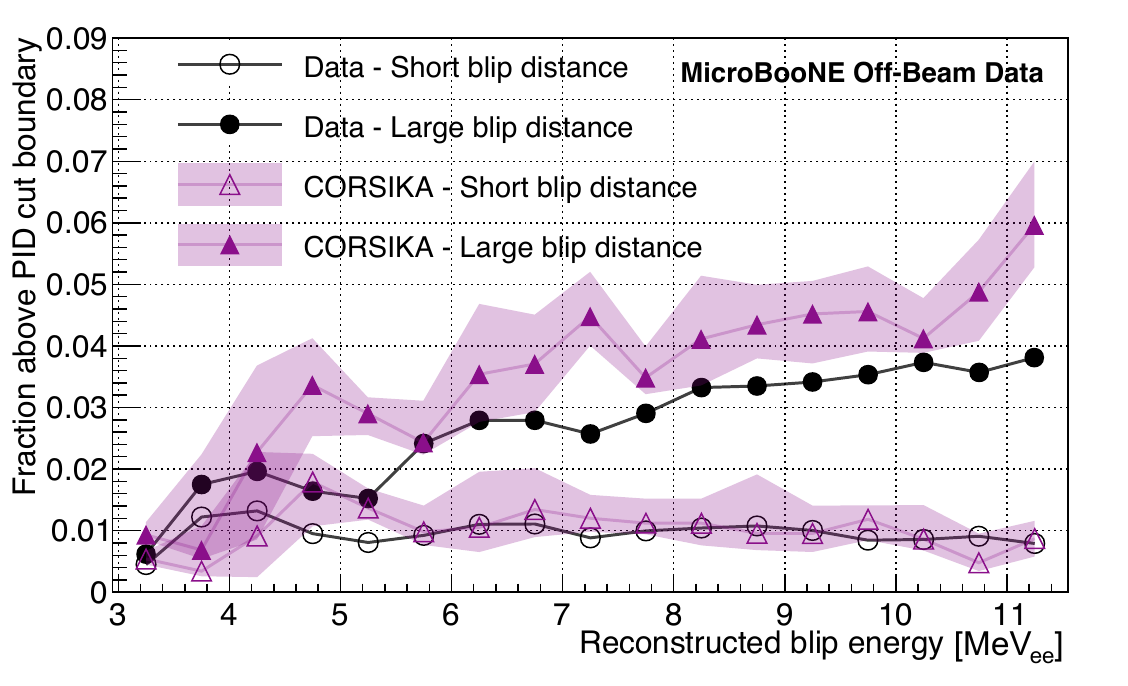}
\caption{Fraction of blips in the short (open markers) and long (solid markers) blip-track proximity categories passing the applied proton-like PID cut.  Shaded bands on MC data points represent systematic uncertainties.  
While the fraction for short proximity blips is roughly 1\% by construction, the fraction for the long blip-track proximity sample deviates strongly from this value, indicating the presence of true protons in this sample. The fractions shown for the CORSIKA MC simulations indicate an over-prediction of cosmogenically-produced protons.}
\label{fig:proton_frac}
\end{figure}

Applying the data-derived PID cut to all blips with energies greater than 3 MeV, we select a substantially higher fraction of proton-like blip candidates in the sample farther from tracks, (2.06$\pm$0.01)\%, compared to that selected in the close-proximity sample, which is $\sim$1\% by construction. This proton-like fraction, plotted as a function of $E_\text{blip}$ in Fig.~\ref{fig:proton_frac}, increases with reconstructed energy, reaching an average of $(3.12 \pm 0.03)$\% above 6 MeV. In regions that are less dominated by tracks and their associated bremsstrahlung radiation within MicroBooNE event readouts, an additional population of highly-ionizing, proton-like energy depositions is clearly visible.

As shown in Fig.~\ref{fig:proton_frac}, a comparatively high fraction of proton-like blips is also visible in the CORSIKA MC's long blip-track proximity subset.  
In the previous subsection, it was established that this higher proton-like blip fraction indeed derived from the presence of true proton energy depositions generated by cosmogenic neutrons.  
This strengthens confidence that the selected proton-like sample is largely composed of true cosmogenically-produced protons.  

\begin{figure}
\includegraphics[width=0.5\textwidth]{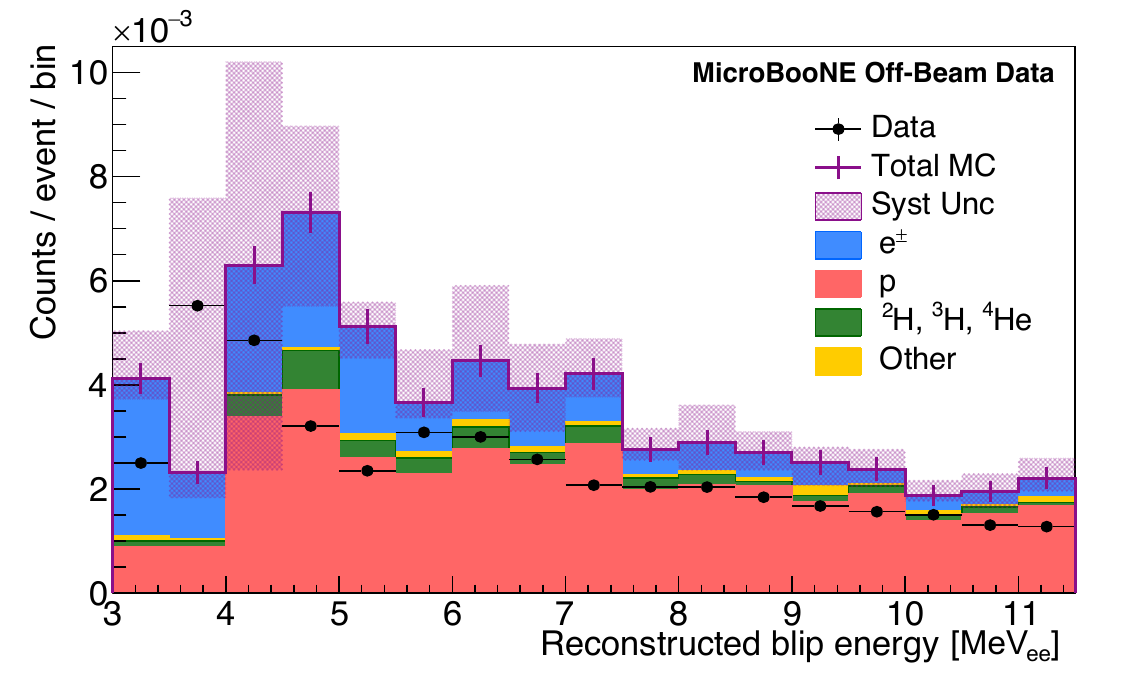}
\caption{Measured (black points) and simulated (stacked histograms) reconstructed energy spectrum of selected proton-like events in MicroBooNE's long blip-track proximity sample. Purple vertical lines represent the statistical uncertainty in the MC dataset.}
\label{fig:proton_en}
\end{figure}

While these cosmogenically-generated proton blips are observable in both data and CORSIKA MC, the overall proportion is substantially lower in data.  
This difference is visible in Fig.~\ref{fig:proton_frac}, as well as in Fig.~\ref{fig:proton_en}, which shows the predicted and measured \Eb spectrum of all selected proton-like blips in the long blip-track proximity subset.  
While spectral features are generally comparable, the overall normalization of the two does not agree.  

Above 3 MeV, a total of 27,717$\pm$166~\text{(stat)} proton-like blips are measured, while simulations predict 39,684$^{+12714} _{ -8073}$~\text{(syst)}$\pm$199~\text{(stat)}, corresponding to 0.0425$\pm$0.0003 proton-like blips per event for data and 0.0607$^{+0.0195}_{-0.0124}$ for MC.   Similarly, above 6 MeV, we expect 
20,842$^{+4057}_{-3140}$~\text{(syst)}$\pm$144~\text{(stat)} counts but we only detect 13,653$\pm$117~\text{(stat)}, that is, 0.0319$^{+0.0062}_{-0.0048}$ and 0.0209$\pm$0.0002 proton-like blips per event in MC and data respectively. Stated in terms of data-MC ratios, above 3~MeV, the ratio of proton-like blips in data compared to MC is only about 0.70$\pm$0.18, with the uncertainty dominated by systematic effects. Above 6~MeV, where our proton purity is expected to be higher, this data-to-MC ratio is 0.66$\pm$0.11. This deficit becomes substantially more pronounced when the MC-derived PID cuts are used on data, rather than the data-driven ones. The significance of the data-MC difference is limited by systematic uncertainties from electron-ion recombination modeling and blip energy scales, with both producing substantial shifts in the fitted PID boundary.

Differences in measured and predicted cosmogenic protons could be due to incorrect predictions of cosmogenic neutron fluxes in the CORSIKA generator, and/or to incorrect neutron transport in the LArTPC and surrounding material by Geant4.  
As demonstrated in previous MicroBooNE cosmic ray studies~\cite{MicroBooNE:2020fmc}, muon fluxes predicted at Earth's surface differ by $\sim$40\% when considering incident cosmic protons versus the full array of nuclear fragments, with the former providing a better fit to MicroBooNE LArTPC track datasets.  
Even in cases where predicted muon rates in MicroBooNE are found to be relatively consistent between the different cosmic ray generator packages CORSIKA and CRY~\cite{Hagmann:2007ziw}, predicted neutron fluxes incident on the TPC are still offset between these generators by more than a factor of 2~\cite{MicroBooNE:2016amq}.  
These past studies lend credence to the notion that incident neutron fluxes may be incorrectly predicted in the CORSIKA simulation used for this study.  
On the other hand, the inclusive interaction cross section of high-energy neutrons on argon nuclei has only recently been measured~\cite{CAPTAIN:2019fxo,CAPTAIN:2022nzf}, and no exclusive measurements of proton-producing interaction channels have been performed to date.  
Thus, it is also reasonable to attribute an excess of predicted cosmogenically-produced protons to mismodeling of cosmic neutron interactions within the MicroBooNE LArTPC by Geant4.  

\subsubsection{General Purpose Proton Selection}
\label{subsec:generic}

The pure sample of low-energy protons described in the previous section was selected with a blip PID metric and cut values tuned for a specific use case: isolating a small and sub-dominant cosmogenically-produced proton population amidst a dominant sea of low-energy electromagnetic cosmogenic backgrounds.  
In this case, an \Eb-independent 99\% electron blip rejection factor was sufficient for achieving better than 50\% purity for true proton signatures above $4~\text{MeV}_\text{ee}$ \Eb, a value above which a substantial fraction of cosmogenically-produced protons reside.  
Other use cases for $p$-$e^\pm$ separation may call for differing levels of background rejection or efficiency than were implemented here.  
In an appendix that accompanies this manuscript we provide more general quantitative depictions and descriptions of blip energy response and low-energy particle discrimination capability for electrons and protons in the MicroBooNE detector.  
This appendix, alongside proton and electron response and PID data files, should enable the community to more broadly explore potential applications of low-energy $p$-$e^\pm$ discrimination in neutrino LArTPCs. While LArTPCs operating at electric fields higher than MicroBooNE’s nominal 274~V/cm may achieve better charge-based detection efficiency for low-energy particles, the information provided with this manuscript offers a baseline level of performance, validated on data, from which future studies can build. 

\section{Summary}
\label{sec:summary}

Using ambient radiogenic and cosmogenic activity in the MicroBooNE detector, we have demonstrated new energy calibration and particle discrimination capabilities at the MeV scale in large neutrino LArTPCs.  
Isolated low-energy activity was identified and reconstructed in MicroBooNE data and MC simulations using a custom-built LArSoft toolkit, \texttt{BlipReco}, validated in previous MicroBooNE results.  
We performed detailed analysis on $>$600,000 MicroBooNE events recorded in the absence of BNB beam activity, with an average of roughly 106 blips observed per event.  

The lowest-energy blips within the reconstructed sample exhibited multiple spectral features, including a prominent shoulder in the 2-3~MeV$_\text{ee}$ \Eb range, which was found to primarily originate from 2.614~MeV $\gamma$~rays released by $^{208}$Tl present in G10 fiberglass mechanical struts that support the MicroBooNE TPC and field cage.  
This feature was used to perform a percent-level precision calibration of blip energy scales in MicroBooNE, with data and MC features found to agree within $(3.1 \pm~0.2 \text{ (stat)} \pm~1.2 \text{ (syst)})\%$.  
This demonstration is vital for performing reliable high-level physics analysis with MeV-scale blips in MicroBooNE.  
It also provides a useful blueprint that can be followed in the calibration of future large LArTPC detectors.  
In particular, the underground DUNE experiment will be unable to perform LArTPC response calibrations with large cosmic muon datasets, and will likely rely on MeV-scale calibration strategies akin to those demonstrated in this paper.  

Study of the 2.614~MeV $^{208}$Tl $\gamma$-ray Compton edge in MicroBooNE also enabled a measurement of the specific activity of $^{208}$Tl in its fiberglass structural materials, which was found to be $(11.7 \pm~0.2~\text{(stat)} \pm~3.1~\text{(syst)})~\text{Bq/kg}$.
This measurement is relevant to other large neutrino LArTPC experiments, since all incorporate fiberglass structural materials in the immediate vicinity of their active TPC volume.  
While this background source has no direct impact on the primary MicroBooNE physics goals, its presence serves as a cautionary tale for future LArTPC efforts with ambitious MeV-scale physics programs.  
In particular, a comparable level of $^{208}$Tl activity in the DUNE far detector may represent a potential challenge for the triggering and reconstruction of astrophysical neutrinos.  
Our result emphasizes the necessity for dedicated material screening campaigns for DUNE and other low-energy single-phase LArTPCs.  
  
New particle discrimination capabilities at the MeV-scale were explored using cosmogenic blips with more than 3~\MeVee of deposited energy.
By making selections on the charge-to-size ratio of reconstructed blips, we identified a purified sample of topologically isolated low-energy protons generated by inelastic scatters of cosmic neutrons.  
With true proton kinetic energies ranging from $\sim$15-40~MeV, the dataset is the lowest-energy of its kind ever identified in a neutrino LArTPC.
While typical MicroBooNE events contain more than a dozen reconstructed cosmic muon tracks, we identified roughly one reconstructed neutron-induced proton blip candidate per two MicroBooNE events.  
This rate of proton-like blip detection was found to be substantially lower than cosmic ray MC predictions, indicating neutron flux or transport issues in MicroBooNE's CORSIKA and Geant4 based cosmic simulation.

This analysis represents the first-ever exploration of particle discrimination capabilities at MeV energy scales in a large neutrino LArTPC, and it opens doors to a range of future novel cross section and BSM physics measurements in these detectors.  
To further aid experimentalists and phenomenologists, we have provided an appendix and data files with descriptions of low-energy response and particle discrimination capabilities for the MicroBooNE LArTPC.  


\begin{acknowledgments}
{
This document was prepared by the MicroBooNE collaboration using the resources of the Fermi National Accelerator Laboratory (Fermilab), a U.S. Department of Energy, Office of Science, HEP User Facility. Fermilab is managed by Fermi Research Alliance, LLC (FRA), acting under Contract No. DE-AC02-07CH11359.  MicroBooNE is supported by the following: the U.S. Department of Energy, Office of Science, Offices of High Energy Physics and Nuclear Physics; the U.S. National Science Foundation; the Swiss National Science Foundation; the Science and Technology Facilities Council (STFC), part of the United Kingdom Research  and Innovation; the Royal Society (United Kingdom); and the UK Research  and Innovation (UKRI) Future Leaders Fellowship. 
Additional support for the laser calibration system and cosmic ray tagger was provided by the Albert Einstein Center for Fundamental Physics, Bern, Switzerland. 
We also acknowledge the contributions of technical and scientific staff to the design, construction, and operation of the MicroBooNE detector as well as the contributions of past collaborators to the development of MicroBooNE analyses, without whom this work would not have been possible. 
For the purpose of open access, the authors have applied a Creative Commons Attribution (CC BY) public copyright license to any Author Accepted Manuscript version arising from this submission.
}
\end{acknowledgments}

\bibliography{refs}{}

\begin{thebibliography}{76}%
\makeatletter
\providecommand \@ifxundefined [1]{%
 \@ifx{#1\undefined}
}%
\providecommand \@ifnum [1]{%
 \ifnum #1\expandafter \@firstoftwo
 \else \expandafter \@secondoftwo
 \fi
}%
\providecommand \@ifx [1]{%
 \ifx #1\expandafter \@firstoftwo
 \else \expandafter \@secondoftwo
 \fi
}%
\providecommand \natexlab [1]{#1}%
\providecommand \enquote  [1]{``#1''}%
\providecommand \bibnamefont  [1]{#1}%
\providecommand \bibfnamefont [1]{#1}%
\providecommand \citenamefont [1]{#1}%
\providecommand \href@noop [0]{\@secondoftwo}%
\providecommand \href [0]{\begingroup \@sanitize@url \@href}%
\providecommand \@href[1]{\@@startlink{#1}\@@href}%
\providecommand \@@href[1]{\endgroup#1\@@endlink}%
\providecommand \@sanitize@url [0]{\catcode `\\12\catcode `\$12\catcode `\&12\catcode `\#12\catcode `\^12\catcode `\_12\catcode `\%12\relax}%
\providecommand \@@startlink[1]{}%
\providecommand \@@endlink[0]{}%
\providecommand \url  [0]{\begingroup\@sanitize@url \@url }%
\providecommand \@url [1]{\endgroup\@href {#1}{\urlprefix }}%
\providecommand \urlprefix  [0]{URL }%
\providecommand \Eprint [0]{\href }%
\providecommand \doibase [0]{https://doi.org/}%
\providecommand \selectlanguage [0]{\@gobble}%
\providecommand \bibinfo  [0]{\@secondoftwo}%
\providecommand \bibfield  [0]{\@secondoftwo}%
\providecommand \translation [1]{[#1]}%
\providecommand \BibitemOpen [0]{}%
\providecommand \bibitemStop [0]{}%
\providecommand \bibitemNoStop [0]{.\EOS\space}%
\providecommand \EOS [0]{\spacefactor3000\relax}%
\providecommand \BibitemShut  [1]{\csname bibitem#1\endcsname}%
\let\auto@bib@innerbib\@empty
\bibitem [{\citenamefont {Anderson}\ \emph {et~al.}(2012)\citenamefont {Anderson} \emph {et~al.}}]{Anderson:2012vc}%
  \BibitemOpen
  \bibfield  {author} {\bibinfo {author} {\bibfnamefont {C.}~\bibnamefont {Anderson}} \emph {et~al.} (\bibinfo {collaboration} {ArgoNeuT}),\ }\bibfield  {title} {\bibinfo {title} {{The ArgoNeuT detector in the NuMI low-energy beam line at Fermilab}},\ }\href {https://doi.org/10.1088/1748-0221/7/10/P10019} {\bibfield  {journal} {\bibinfo  {journal} {{J. Instrum.}}\ }\textbf {\bibinfo {volume} {7}},\ \bibinfo {pages} {P10019} (\bibinfo {year} {2012})}\BibitemShut {NoStop}%
\bibitem [{\citenamefont {Acciarri}\ \emph {et~al.}(2017{\natexlab{a}})\citenamefont {Acciarri} \emph {et~al.}}]{ub_det}%
  \BibitemOpen
  \bibfield  {author} {\bibinfo {author} {\bibfnamefont {R.}~\bibnamefont {Acciarri}} \emph {et~al.} (\bibinfo {collaboration} {MicroBooNE}),\ }\bibfield  {title} {\bibinfo {title} {{Design and construction of the MicroBooNE detector}},\ }\href {https://doi.org/10.1088/1748-0221/12/02/P02017} {\bibfield  {journal} {\bibinfo  {journal} {{J. Instrum.}}\ }\textbf {\bibinfo {volume} {12}},\ \bibinfo {pages} {P02017} (\bibinfo {year} {2017}{\natexlab{a}})}\BibitemShut {NoStop}%
\bibitem [{\citenamefont {Acciarri}\ \emph {et~al.}(2020{\natexlab{a}})\citenamefont {Acciarri} \emph {et~al.}}]{lariat_detpaper}%
  \BibitemOpen
  \bibfield  {author} {\bibinfo {author} {\bibfnamefont {R.}~\bibnamefont {Acciarri}} \emph {et~al.} (\bibinfo {collaboration} {LArIAT}),\ }\bibfield  {title} {\bibinfo {title} {{The Liquid Argon In A Testbeam (LArIAT) Experiment}},\ }\href {https://doi.org/10.1088/1748-0221/15/04/P04026} {\bibfield  {journal} {\bibinfo  {journal} {{J. Instrum.}}\ }\textbf {\bibinfo {volume} {15}},\ \bibinfo {pages} {P04026} (\bibinfo {year} {2020}{\natexlab{a}})}\BibitemShut {NoStop}%
\bibitem [{\citenamefont {Abratenko}\ \emph {et~al.}(2023{\natexlab{a}})\citenamefont {Abratenko} \emph {et~al.}}]{ICARUS:2023gpo}%
  \BibitemOpen
  \bibfield  {author} {\bibinfo {author} {\bibfnamefont {P.}~\bibnamefont {Abratenko}} \emph {et~al.} (\bibinfo {collaboration} {ICARUS}),\ }\bibfield  {title} {\bibinfo {title} {{ICARUS at the Fermilab Short-Baseline Neutrino program: initial operation}},\ }\href {https://doi.org/10.1140/epjc/s10052-023-11610-y} {\bibfield  {journal} {\bibinfo  {journal} {Eur. Phys. J. C}\ }\textbf {\bibinfo {volume} {83}},\ \bibinfo {pages} {467} (\bibinfo {year} {2023}{\natexlab{a}})}\BibitemShut {NoStop}%
\bibitem [{\citenamefont {Acciarri}\ \emph {et~al.}(2017{\natexlab{b}})\citenamefont {Acciarri} \emph {et~al.}}]{MicroBooNE:2017qiu}%
  \BibitemOpen
  \bibfield  {author} {\bibinfo {author} {\bibfnamefont {R.}~\bibnamefont {Acciarri}} \emph {et~al.} (\bibinfo {collaboration} {MicroBooNE}),\ }\bibfield  {title} {\bibinfo {title} {{Noise Characterization and Filtering in the MicroBooNE Liquid Argon TPC}},\ }\href {https://doi.org/10.1088/1748-0221/12/08/P08003} {\bibfield  {journal} {\bibinfo  {journal} {{J. Instrum.}}\ }\textbf {\bibinfo {volume} {12}},\ \bibinfo {pages} {P08003} (\bibinfo {year} {2017}{\natexlab{b}})}\BibitemShut {NoStop}%
\bibitem [{\citenamefont {Acciarri}\ \emph {et~al.}(2019)\citenamefont {Acciarri} \emph {et~al.}}]{argo_mev}%
  \BibitemOpen
  \bibfield  {author} {\bibinfo {author} {\bibfnamefont {R.}~\bibnamefont {Acciarri}} \emph {et~al.} (\bibinfo {collaboration} {ArgoNeuT}),\ }\bibfield  {title} {\bibinfo {title} {{Demonstration of MeV-scale physics in liquid argon time projection chambers using ArgoNeuT}},\ }\href {https://doi.org/10.1103/PhysRevD.99.012002} {\bibfield  {journal} {\bibinfo  {journal} {Phys. Rev. D}\ }\textbf {\bibinfo {volume} {99}},\ \bibinfo {pages} {012002} (\bibinfo {year} {2019})}\BibitemShut {NoStop}%
\bibitem [{\citenamefont {Abratenko}\ \emph {et~al.}(2024{\natexlab{a}})\citenamefont {Abratenko} \emph {et~al.}}]{MicroBooNE:2024hun}%
  \BibitemOpen
  \bibfield  {author} {\bibinfo {author} {\bibfnamefont {P.}~\bibnamefont {Abratenko}} \emph {et~al.} (\bibinfo {collaboration} {MicroBooNE}),\ }\bibfield  {title} {\bibinfo {title} {{Demonstration of neutron identification in neutrino interactions in the MicroBooNE liquid argon time projection chamber}},\ }\href {https://doi.org/https://doi.org/10.1140/epjc/s10052-024-13423-z} {\bibfield  {journal} {\bibinfo  {journal} {Eur. Phys. J. C}\ }\textbf {\bibinfo {volume} {84}},\ \bibinfo {pages} {1052} (\bibinfo {year} {2024}{\natexlab{a}})}\BibitemShut {NoStop}%
\bibitem [{\citenamefont {Rivera}(2022)}]{Rivera:2021dcf}%
  \BibitemOpen
  \bibfield  {author} {\bibinfo {author} {\bibfnamefont {D.~O.}\ \bibnamefont {Rivera}},\ }\emph {\bibinfo {title} {{Neutron Cross Section Measurement In The ProtoDUNE-SP Experiment}}},\ \href@noop {} {Ph.D. thesis},\ \bibinfo  {school} {UPenn, Philadelphia, Pennsylvania U.} (\bibinfo {year} {2022}),\ \bibinfo {note} {{\href{https://www.osti.gov/biblio/1836573}{FERMILAB-THESIS-2021-31}}}\BibitemShut {NoStop}%
\bibitem [{\citenamefont {Ankowski}\ \emph {et~al.}(2015)\citenamefont {Ankowski}, \citenamefont {Coloma}, \citenamefont {Huber}, \citenamefont {Mariani},\ and\ \citenamefont {Vagnoni}}]{Ankowski:2015kya}%
  \BibitemOpen
  \bibfield  {author} {\bibinfo {author} {\bibfnamefont {A.~M.}\ \bibnamefont {Ankowski}}, \bibinfo {author} {\bibfnamefont {P.}~\bibnamefont {Coloma}}, \bibinfo {author} {\bibfnamefont {P.}~\bibnamefont {Huber}}, \bibinfo {author} {\bibfnamefont {C.}~\bibnamefont {Mariani}},\ and\ \bibinfo {author} {\bibfnamefont {E.}~\bibnamefont {Vagnoni}},\ }\bibfield  {title} {\bibinfo {title} {{Missing energy and the measurement of the CP-violating phase in neutrino oscillations}},\ }\href {https://doi.org/10.1103/PhysRevD.92.091301} {\bibfield  {journal} {\bibinfo  {journal} {Phys. Rev. D}\ }\textbf {\bibinfo {volume} {92}},\ \bibinfo {pages} {091301} (\bibinfo {year} {2015})}\BibitemShut {NoStop}%
\bibitem [{\citenamefont {Friedland}\ and\ \citenamefont {Li}(2019)}]{Friedland:2018vry}%
  \BibitemOpen
  \bibfield  {author} {\bibinfo {author} {\bibfnamefont {A.}~\bibnamefont {Friedland}}\ and\ \bibinfo {author} {\bibfnamefont {S.~W.}\ \bibnamefont {Li}},\ }\bibfield  {title} {\bibinfo {title} {{Understanding the energy resolution of liquid argon neutrino detectors}},\ }\href {https://doi.org/10.1103/PhysRevD.99.036009} {\bibfield  {journal} {\bibinfo  {journal} {Phys. Rev. D}\ }\textbf {\bibinfo {volume} {99}},\ \bibinfo {pages} {036009} (\bibinfo {year} {2019})}\BibitemShut {NoStop}%
\bibitem [{\citenamefont {Abi}\ \emph {et~al.}(2020)\citenamefont {Abi} \emph {et~al.}}]{DUNE:2020txw}%
  \BibitemOpen
  \bibfield  {author} {\bibinfo {author} {\bibfnamefont {B.}~\bibnamefont {Abi}} \emph {et~al.} (\bibinfo {collaboration} {DUNE}),\ }\bibfield  {title} {\bibinfo {title} {{Deep Underground Neutrino Experiment (DUNE), Far Detector Technical Design Report, Volume IV: Far Detector Single-phase Technology}},\ }\href {https://doi.org/10.1088/1748-0221/15/08/T08010} {\bibfield  {journal} {\bibinfo  {journal} {{J. Instrum.}}\ }\textbf {\bibinfo {volume} {15}},\ \bibinfo {pages} {T08010} (\bibinfo {year} {2020})}\BibitemShut {NoStop}%
\bibitem [{\citenamefont {Aguilar-Arevalo}\ \emph {et~al.}(2021)\citenamefont {Aguilar-Arevalo} \emph {et~al.}}]{MiniBooNE:2020pnu}%
  \BibitemOpen
  \bibfield  {author} {\bibinfo {author} {\bibfnamefont {A.~A.}\ \bibnamefont {Aguilar-Arevalo}} \emph {et~al.} (\bibinfo {collaboration} {MiniBooNE}),\ }\bibfield  {title} {\bibinfo {title} {{Updated MiniBooNE neutrino oscillation results with increased data and new background studies}},\ }\href {https://doi.org/10.1103/PhysRevD.103.052002} {\bibfield  {journal} {\bibinfo  {journal} {Phys. Rev. D}\ }\textbf {\bibinfo {volume} {103}},\ \bibinfo {pages} {052002} (\bibinfo {year} {2021})}\BibitemShut {NoStop}%
\bibitem [{\citenamefont {Abratenko}\ \emph {et~al.}(2022{\natexlab{a}})\citenamefont {Abratenko} \emph {et~al.}}]{MicroBooNE:2021zai}%
  \BibitemOpen
  \bibfield  {author} {\bibinfo {author} {\bibfnamefont {P.}~\bibnamefont {Abratenko}} \emph {et~al.} (\bibinfo {collaboration} {MicroBooNE}),\ }\bibfield  {title} {\bibinfo {title} {{Search for Neutrino-Induced Neutral-Current \ensuremath{\Delta} Radiative Decay in MicroBooNE and a First Test of the MiniBooNE Low Energy Excess under a Single-Photon Hypothesis}},\ }\href {https://doi.org/10.1103/PhysRevLett.128.111801} {\bibfield  {journal} {\bibinfo  {journal} {Phys. Rev. Lett.}\ }\textbf {\bibinfo {volume} {128}},\ \bibinfo {pages} {111801} (\bibinfo {year} {2022}{\natexlab{a}})}\BibitemShut {NoStop}%
\bibitem [{\citenamefont {Abratenko}\ \emph {et~al.}(2022{\natexlab{b}})\citenamefont {Abratenko} \emph {et~al.}}]{MicroBooNE:2021nxr}%
  \BibitemOpen
  \bibfield  {author} {\bibinfo {author} {\bibfnamefont {P.}~\bibnamefont {Abratenko}} \emph {et~al.} (\bibinfo {collaboration} {MicroBooNE}),\ }\bibfield  {title} {\bibinfo {title} {{Search for an anomalous excess of inclusive charged-current $\nu_e$ interactions in the MicroBooNE experiment using Wire-Cell reconstruction}},\ }\href {https://doi.org/10.1103/PhysRevD.105.112005} {\bibfield  {journal} {\bibinfo  {journal} {Phys. Rev. D}\ }\textbf {\bibinfo {volume} {105}},\ \bibinfo {pages} {112005} (\bibinfo {year} {2022}{\natexlab{b}})}\BibitemShut {NoStop}%
\bibitem [{\citenamefont {Abratenko}\ \emph {et~al.}(2022{\natexlab{c}})\citenamefont {Abratenko} \emph {et~al.}}]{MicroBooNE:2021wad}%
  \BibitemOpen
  \bibfield  {author} {\bibinfo {author} {\bibfnamefont {P.}~\bibnamefont {Abratenko}} \emph {et~al.} (\bibinfo {collaboration} {MicroBooNE}),\ }\bibfield  {title} {\bibinfo {title} {{Search for an anomalous excess of charged-current \ensuremath{\nu_e} interactions without pions in the final state with the MicroBooNE experiment}},\ }\href {https://doi.org/10.1103/PhysRevD.105.112004} {\bibfield  {journal} {\bibinfo  {journal} {Phys. Rev. D}\ }\textbf {\bibinfo {volume} {105}},\ \bibinfo {pages} {112004} (\bibinfo {year} {2022}{\natexlab{c}})}\BibitemShut {NoStop}%
\bibitem [{\citenamefont {Abratenko}\ \emph {et~al.}(2022{\natexlab{d}})\citenamefont {Abratenko} \emph {et~al.}}]{MicroBooNE:2022tdd}%
  \BibitemOpen
  \bibfield  {author} {\bibinfo {author} {\bibfnamefont {P.}~\bibnamefont {Abratenko}} \emph {et~al.} (\bibinfo {collaboration} {MicroBooNE}),\ }\bibfield  {title} {\bibinfo {title} {{Differential cross section measurement of charged current \ensuremath{\nu_e} interactions without final-state pions in MicroBooNE}},\ }\href {https://doi.org/10.1103/PhysRevD.106.L051102} {\bibfield  {journal} {\bibinfo  {journal} {Phys. Rev. D}\ }\textbf {\bibinfo {volume} {106}},\ \bibinfo {pages} {L051102} (\bibinfo {year} {2022}{\natexlab{d}})}\BibitemShut {NoStop}%
\bibitem [{\citenamefont {Abratenko}\ \emph {et~al.}(2023{\natexlab{b}})\citenamefont {Abratenko} \emph {et~al.}}]{MicroBooNE:2022zhr}%
  \BibitemOpen
  \bibfield  {author} {\bibinfo {author} {\bibfnamefont {P.}~\bibnamefont {Abratenko}} \emph {et~al.} (\bibinfo {collaboration} {MicroBooNE}),\ }\bibfield  {title} {\bibinfo {title} {{Measurement of neutral current single \ensuremath{\pi^0} production on argon with the MicroBooNE detector}},\ }\href {https://doi.org/10.1103/PhysRevD.107.012004} {\bibfield  {journal} {\bibinfo  {journal} {Phys. Rev. D}\ }\textbf {\bibinfo {volume} {107}},\ \bibinfo {pages} {012004} (\bibinfo {year} {2023}{\natexlab{b}})}\BibitemShut {NoStop}%
\bibitem [{\citenamefont {Abratenko}\ \emph {et~al.}(2024{\natexlab{b}})\citenamefont {Abratenko} \emph {et~al.}}]{MicroBooNE:2024xod}%
  \BibitemOpen
  \bibfield  {author} {\bibinfo {author} {\bibfnamefont {P.}~\bibnamefont {Abratenko}} \emph {et~al.} (\bibinfo {collaboration} {MicroBooNE Collaboration}),\ }\bibfield  {title} {\bibinfo {title} {Inclusive cross section measurements in final states with and without protons for charged-current ${\ensuremath{\nu}}_{\ensuremath{\mu}}$-ar scattering in microboone},\ }\href {https://doi.org/10.1103/PhysRevD.110.013006} {\bibfield  {journal} {\bibinfo  {journal} {Phys. Rev. D}\ }\textbf {\bibinfo {volume} {110}},\ \bibinfo {pages} {013006} (\bibinfo {year} {2024}{\natexlab{b}})}\BibitemShut {NoStop}%
\bibitem [{\citenamefont {Abratenko}\ \emph {et~al.}(2024{\natexlab{c}})\citenamefont {Abratenko} \emph {et~al.}}]{MicroBooNE:2024klj}%
  \BibitemOpen
  \bibfield  {author} {\bibinfo {author} {\bibfnamefont {P.}~\bibnamefont {Abratenko}} \emph {et~al.} (\bibinfo {collaboration} {MicroBooNE Collaboration}),\ }\bibfield  {title} {\bibinfo {title} {First simultaneous measurement of differential muon-neutrino charged-current cross sections on argon for final states with and without protons using microboone data},\ }\href {https://doi.org/10.1103/PhysRevLett.133.041801} {\bibfield  {journal} {\bibinfo  {journal} {Phys. Rev. Lett.}\ }\textbf {\bibinfo {volume} {133}},\ \bibinfo {pages} {041801} (\bibinfo {year} {2024}{\natexlab{c}})}\BibitemShut {NoStop}%
\bibitem [{\citenamefont {Abratenko}\ \emph {et~al.}(2024{\natexlab{d}})\citenamefont {Abratenko} \emph {et~al.}}]{MicroBooNE:2024tmp}%
  \BibitemOpen
  \bibfield  {author} {\bibinfo {author} {\bibfnamefont {P.}~\bibnamefont {Abratenko}} \emph {et~al.} (\bibinfo {collaboration} {MicroBooNE}),\ }\bibfield  {title} {\bibinfo {title} {{Measurement of double-differential cross sections for mesonless charged-current muon neutrino interactions on argon with final-state protons using the MicroBooNE detector}},\ }\href@noop {} {\  (\bibinfo {year} {2024}{\natexlab{d}})},\ \Eprint {https://arxiv.org/abs/2403.19574} {arXiv:2403.19574 [hep-ex]} \BibitemShut {NoStop}%
\bibitem [{\citenamefont {Benevides~Rodrigues}(2022)}]{BenevidesRodrigues:2022wxz}%
  \BibitemOpen
  \bibfield  {author} {\bibinfo {author} {\bibfnamefont {O.}~\bibnamefont {Benevides~Rodrigues}},\ }\emph {\bibinfo {title} {{Search for NuMI $\mu$DAR Electron Neutrinos in MicroBooNE}}},\ \href@noop {} {Ph.D. thesis},\ \bibinfo  {school} {Syracuse U.} (\bibinfo {year} {2022}),\ \bibinfo {note} {{\href{https://www.osti.gov/biblio/1906068}{FERMILAB-THESIS-2022-19}}}\BibitemShut {NoStop}%
\bibitem [{\citenamefont {Grant}\ and\ \citenamefont {Littlejohn}(2016)}]{Grant:2015jva}%
  \BibitemOpen
  \bibfield  {author} {\bibinfo {author} {\bibfnamefont {C.}~\bibnamefont {Grant}}\ and\ \bibinfo {author} {\bibfnamefont {B.}~\bibnamefont {Littlejohn}},\ }\bibfield  {title} {\bibinfo {title} {{Opportunities With Decay-At-Rest Neutrinos From Decay-In-Flight Neutrino Beams}},\ }\href {https://doi.org/10.22323/1.282.0483} {\bibfield  {journal} {\bibinfo  {journal} {PoS}\ }\textbf {\bibinfo {volume} {ICHEP2016}},\ \bibinfo {pages} {483} (\bibinfo {year} {2016})}\BibitemShut {NoStop}%
\bibitem [{\citenamefont {Harnik}\ \emph {et~al.}(2020)\citenamefont {Harnik}, \citenamefont {Kelly},\ and\ \citenamefont {Machado}}]{Harnik:2019iwv}%
  \BibitemOpen
  \bibfield  {author} {\bibinfo {author} {\bibfnamefont {R.}~\bibnamefont {Harnik}}, \bibinfo {author} {\bibfnamefont {K.~J.}\ \bibnamefont {Kelly}},\ and\ \bibinfo {author} {\bibfnamefont {P.~A.~N.}\ \bibnamefont {Machado}},\ }\bibfield  {title} {\bibinfo {title} {{Prospects of Measuring Oscillated Decay-at-Rest Neutrinos at Long Baselines}},\ }\href {https://doi.org/10.1103/PhysRevD.101.033008} {\bibfield  {journal} {\bibinfo  {journal} {Phys. Rev. D}\ }\textbf {\bibinfo {volume} {101}},\ \bibinfo {pages} {033008} (\bibinfo {year} {2020})}\BibitemShut {NoStop}%
\bibitem [{\citenamefont {Castiglioni}\ \emph {et~al.}(2020)\citenamefont {Castiglioni}, \citenamefont {Foreman}, \citenamefont {Lepetic}, \citenamefont {Littlejohn}, \citenamefont {Malaker},\ and\ \citenamefont {Mastbaum}}]{Castiglioni:2020tsu}%
  \BibitemOpen
  \bibfield  {author} {\bibinfo {author} {\bibfnamefont {W.}~\bibnamefont {Castiglioni}}, \bibinfo {author} {\bibfnamefont {W.}~\bibnamefont {Foreman}}, \bibinfo {author} {\bibfnamefont {I.}~\bibnamefont {Lepetic}}, \bibinfo {author} {\bibfnamefont {B.~R.}\ \bibnamefont {Littlejohn}}, \bibinfo {author} {\bibfnamefont {M.}~\bibnamefont {Malaker}},\ and\ \bibinfo {author} {\bibfnamefont {A.}~\bibnamefont {Mastbaum}},\ }\bibfield  {title} {\bibinfo {title} {{Benefits of MeV-scale reconstruction capabilities in large liquid argon time projection chambers}},\ }\href {https://doi.org/10.1103/PhysRevD.102.092010} {\bibfield  {journal} {\bibinfo  {journal} {Phys. Rev. D}\ }\textbf {\bibinfo {volume} {102}},\ \bibinfo {pages} {092010} (\bibinfo {year} {2020})}\BibitemShut {NoStop}%
\bibitem [{\citenamefont {Machado}\ \emph {et~al.}(2019)\citenamefont {Machado}, \citenamefont {Palamara},\ and\ \citenamefont {Schmitz}}]{sbnd_phys}%
  \BibitemOpen
  \bibfield  {author} {\bibinfo {author} {\bibfnamefont {P.~A.}\ \bibnamefont {Machado}}, \bibinfo {author} {\bibfnamefont {O.}~\bibnamefont {Palamara}},\ and\ \bibinfo {author} {\bibfnamefont {D.~W.}\ \bibnamefont {Schmitz}},\ }\bibfield  {title} {\bibinfo {title} {{The Short-Baseline Neutrino Program at Fermilab}},\ }\href {https://doi.org/10.1146/annurev-nucl-101917-020949} {\bibfield  {journal} {\bibinfo  {journal} {Ann. Rev. Nucl. Part. Sci.}\ }\textbf {\bibinfo {volume} {69}},\ \bibinfo {pages} {363} (\bibinfo {year} {2019})}\BibitemShut {NoStop}%
\bibitem [{\citenamefont {Andringa}\ \emph {et~al.}(2023)\citenamefont {Andringa} \emph {et~al.}}]{leplar_paper}%
  \BibitemOpen
  \bibfield  {author} {\bibinfo {author} {\bibfnamefont {S.}~\bibnamefont {Andringa}} \emph {et~al.},\ }\bibfield  {title} {\bibinfo {title} {{Low-energy physics in neutrino LArTPCs}},\ }\href {https://doi.org/10.1088/1361-6471/acad17} {\bibfield  {journal} {\bibinfo  {journal} {J. Phys. G}\ }\textbf {\bibinfo {volume} {50}},\ \bibinfo {pages} {033001} (\bibinfo {year} {2023})}\BibitemShut {NoStop}%
\bibitem [{\citenamefont {Abi}\ \emph {et~al.}(2021)\citenamefont {Abi} \emph {et~al.}}]{DUNE:2020zfm}%
  \BibitemOpen
  \bibfield  {author} {\bibinfo {author} {\bibfnamefont {B.}~\bibnamefont {Abi}} \emph {et~al.} (\bibinfo {collaboration} {DUNE}),\ }\bibfield  {title} {\bibinfo {title} {{Supernova neutrino burst detection with the Deep Underground Neutrino Experiment}},\ }\href {https://doi.org/10.1140/epjc/s10052-021-09166-w} {\bibfield  {journal} {\bibinfo  {journal} {Eur. Phys. J. C}\ }\textbf {\bibinfo {volume} {81}},\ \bibinfo {pages} {423} (\bibinfo {year} {2021})}\BibitemShut {NoStop}%
\bibitem [{\citenamefont {Kubota}\ \emph {et~al.}(2022)\citenamefont {Kubota} \emph {et~al.}}]{Q-Pix:2022zjm}%
  \BibitemOpen
  \bibfield  {author} {\bibinfo {author} {\bibfnamefont {S.}~\bibnamefont {Kubota}} \emph {et~al.} (\bibinfo {collaboration} {Q-Pix}),\ }\bibfield  {title} {\bibinfo {title} {{Enhanced low-energy supernova burst detection in large liquid argon time projection chambers enabled by Q-Pix}},\ }\href {https://doi.org/10.1103/PhysRevD.106.032011} {\bibfield  {journal} {\bibinfo  {journal} {Phys. Rev. D}\ }\textbf {\bibinfo {volume} {106}},\ \bibinfo {pages} {032011} (\bibinfo {year} {2022})}\BibitemShut {NoStop}%
\bibitem [{\citenamefont {Capozzi}\ \emph {et~al.}(2019)\citenamefont {Capozzi}, \citenamefont {Li}, \citenamefont {Zhu},\ and\ \citenamefont {Beacom}}]{dune_solar}%
  \BibitemOpen
  \bibfield  {author} {\bibinfo {author} {\bibfnamefont {F.}~\bibnamefont {Capozzi}}, \bibinfo {author} {\bibfnamefont {S.~W.}\ \bibnamefont {Li}}, \bibinfo {author} {\bibfnamefont {G.}~\bibnamefont {Zhu}},\ and\ \bibinfo {author} {\bibfnamefont {J.~F.}\ \bibnamefont {Beacom}},\ }\bibfield  {title} {\bibinfo {title} {{DUNE as the Next-Generation Solar Neutrino Experiment}},\ }\href {https://doi.org/10.1103/PhysRevLett.123.131803} {\bibfield  {journal} {\bibinfo  {journal} {Phys. Rev. Lett.}\ }\textbf {\bibinfo {volume} {123}},\ \bibinfo {pages} {131803} (\bibinfo {year} {2019})}\BibitemShut {NoStop}%
\bibitem [{\citenamefont {Amoruso}\ \emph {et~al.}(2004)\citenamefont {Amoruso} \emph {et~al.}}]{icarus_michel}%
  \BibitemOpen
  \bibfield  {author} {\bibinfo {author} {\bibfnamefont {S.}~\bibnamefont {Amoruso}} \emph {et~al.} (\bibinfo {collaboration} {ICARUS}),\ }\bibfield  {title} {\bibinfo {title} {{Measurement of the $\mu$ decay spectrum with the ICARUS liquid argon TPC}},\ }\href {https://doi.org/10.1140/epjc/s2004-01597-7} {\bibfield  {journal} {\bibinfo  {journal} {Eur. Phys. J. C}\ }\textbf {\bibinfo {volume} {33}},\ \bibinfo {pages} {233} (\bibinfo {year} {2004})}\BibitemShut {NoStop}%
\bibitem [{\citenamefont {Acciarri}\ \emph {et~al.}(2017{\natexlab{c}})\citenamefont {Acciarri} \emph {et~al.}}]{ub_michel}%
  \BibitemOpen
  \bibfield  {author} {\bibinfo {author} {\bibfnamefont {R.}~\bibnamefont {Acciarri}} \emph {et~al.} (\bibinfo {collaboration} {MicroBooNE}),\ }\bibfield  {title} {\bibinfo {title} {{Michel electron reconstruction using cosmic-ray data from the MicroBooNE LArTPC}},\ }\href {https://doi.org/10.1088/1748-0221/12/09/P09014} {\bibfield  {journal} {\bibinfo  {journal} {{J. Instrum.}}\ }\textbf {\bibinfo {volume} {12}},\ \bibinfo {pages} {P09014} (\bibinfo {year} {2017}{\natexlab{c}})}\BibitemShut {NoStop}%
\bibitem [{\citenamefont {Acciarri}\ \emph {et~al.}(2020{\natexlab{b}})\citenamefont {Acciarri} \emph {et~al.}}]{argo_mcp}%
  \BibitemOpen
  \bibfield  {author} {\bibinfo {author} {\bibfnamefont {R.}~\bibnamefont {Acciarri}} \emph {et~al.} (\bibinfo {collaboration} {ArgoNeuT}),\ }\bibfield  {title} {\bibinfo {title} {{Improved Limits on Millicharged Particles Using the ArgoNeuT Experiment at Fermilab}},\ }\href {https://doi.org/10.1103/PhysRevLett.124.131801} {\bibfield  {journal} {\bibinfo  {journal} {Phys. Rev. Lett.}\ }\textbf {\bibinfo {volume} {124}},\ \bibinfo {pages} {131801} (\bibinfo {year} {2020}{\natexlab{b}})}\BibitemShut {NoStop}%
\bibitem [{\citenamefont {Bhat}(2021)}]{ub_mev}%
  \BibitemOpen
  \bibfield  {author} {\bibinfo {author} {\bibfnamefont {A.}~\bibnamefont {Bhat}},\ }\emph {\bibinfo {title} {{MeV Scale Physics in MicroBooNE}}},\ \href@noop {} {Ph.D. thesis},\ \bibinfo  {school} {Syracuse University} (\bibinfo {year} {2021}),\ \bibinfo {note} {{\href{https://www.osti.gov/biblio/1824656}{FERMILAB-THESIS-2021-14}}}\BibitemShut {NoStop}%
\bibitem [{\citenamefont {{MicroBooNE Collaboration}}(2018)}]{ub_ar39}%
  \BibitemOpen
  \bibfield  {author} {\bibinfo {author} {\bibnamefont {{MicroBooNE Collaboration}}},\ }\bibfield  {title} {\bibinfo {title} {{Study of Reconstructed $^{39}$Ar Beta Decays at the MicroBooNE Detector}},\ }\href {https://www.osti.gov/biblio/1573057} {\bibfield  {journal} {\bibinfo  {journal} {{MICROBOONE-NOTE-1050-PUB}}\ } (\bibinfo {year} {{2018}})}\BibitemShut {NoStop}%
\bibitem [{\citenamefont {{MicroBooNE Collaboration}}(2024)}]{ub_mev_pub_note}%
  \BibitemOpen
  \bibfield  {author} {\bibinfo {author} {\bibnamefont {{MicroBooNE Collaboration}}},\ }\bibfield  {title} {\bibinfo {title} {Mev-scale physics in microboone},\ }\href {https://www.osti.gov/biblio/2397303} {\bibfield  {journal} {\bibinfo  {journal} {MICROBOONE-NOTE 1076-PUB}\ } (\bibinfo {year} {2024})}\BibitemShut {NoStop}%
\bibitem [{\citenamefont {Abratenko}(2022)}]{ub_radon}%
  \BibitemOpen
  \bibfield  {author} {\bibinfo {author} {\bibfnamefont {P.}~\bibnamefont {Abratenko}} (\bibinfo {collaboration} {MicroBooNE}),\ }\bibfield  {title} {\bibinfo {title} {{Observation of radon mitigation in MicroBooNE by a liquid argon filtration system}},\ }\href {https://doi.org/10.1088/1748-0221/17/11/P11022} {\bibfield  {journal} {\bibinfo  {journal} {{J. Instrum.}}\ }\textbf {\bibinfo {volume} {17}},\ \bibinfo {pages} {P11022} (\bibinfo {year} {2022})}\BibitemShut {NoStop}%
\bibitem [{\citenamefont {Abratenko}\ \emph {et~al.}(2024{\natexlab{e}})\citenamefont {Abratenko} \emph {et~al.}}]{MicroBooNE:2023ftv}%
  \BibitemOpen
  \bibfield  {author} {\bibinfo {author} {\bibfnamefont {P.}~\bibnamefont {Abratenko}} \emph {et~al.} (\bibinfo {collaboration} {MicroBooNE}),\ }\bibfield  {title} {\bibinfo {title} {{Measurement of ambient radon progeny decay rates and energy spectra in liquid argon using the MicroBooNE detector}},\ }\href {https://doi.org/10.1103/PhysRevD.109.052007} {\bibfield  {journal} {\bibinfo  {journal} {Phys. Rev. D}\ }\textbf {\bibinfo {volume} {109}},\ \bibinfo {pages} {052007} (\bibinfo {year} {2024}{\natexlab{e}})}\BibitemShut {NoStop}%
\bibitem [{\citenamefont {Hernandez-Morquecho}\ \emph {et~al.}(2024)\citenamefont {Hernandez-Morquecho} \emph {et~al.}}]{LArIAT:2024otd}%
  \BibitemOpen
  \bibfield  {author} {\bibinfo {author} {\bibfnamefont {M.~A.}\ \bibnamefont {Hernandez-Morquecho}} \emph {et~al.} (\bibinfo {collaboration} {LArIAT}),\ }\bibfield  {title} {\bibinfo {title} {{Measurements of Pion and Muon Nuclear Capture at Rest on Argon in the LArIAT Experiment}},\ }\href@noop {} {\  (\bibinfo {year} {2024})},\ \Eprint {https://arxiv.org/abs/2408.05133} {arXiv:2408.05133 [hep-ex]} \BibitemShut {NoStop}%
\bibitem [{\citenamefont {Foreman}\ \emph {et~al.}(2020)\citenamefont {Foreman} \emph {et~al.}}]{lariat_michels}%
  \BibitemOpen
  \bibfield  {author} {\bibinfo {author} {\bibfnamefont {W.}~\bibnamefont {Foreman}} \emph {et~al.} (\bibinfo {collaboration} {LArIAT}),\ }\bibfield  {title} {\bibinfo {title} {{Calorimetry for low-energy electrons using charge and light in liquid argon}},\ }\href {https://doi.org/10.1103/PhysRevD.101.012010} {\bibfield  {journal} {\bibinfo  {journal} {Phys. Rev. D}\ }\textbf {\bibinfo {volume} {101}},\ \bibinfo {pages} {012010} (\bibinfo {year} {2020})}\BibitemShut {NoStop}%
\bibitem [{\citenamefont {{{The Gund Company}}}()}]{G10}%
  \BibitemOpen
  \bibfield  {author} {\bibinfo {author} {\bibnamefont {{{The Gund Company}}}},\ }\href@noop {} {\bibinfo {title} {{NEMA Grade G-10 Glass Epoxy Laminate}}},\ \bibinfo {note} {\href{https://thegundcompany.com/wp-content/uploads/2016/11/NEMA-G10-EPGC-201-from-The-Gund-Co.pdf}{https://thegundcompany.com/wp-content/uploads/2016/11/NEMA-G10-EPGC-201-from-The-Gund-Co.pdf}}\BibitemShut {NoStop}%
\bibitem [{\citenamefont {Adams}\ \emph {et~al.}(2018{\natexlab{a}})\citenamefont {Adams} \emph {et~al.}}]{MicroBooNE:2018swd}%
  \BibitemOpen
  \bibfield  {author} {\bibinfo {author} {\bibfnamefont {C.}~\bibnamefont {Adams}} \emph {et~al.} (\bibinfo {collaboration} {{MicroBooNE}}),\ }\bibfield  {title} {\bibinfo {title} {{Ionization electron signal processing in single phase LArTPCs. Part I. Algorithm Description and quantitative evaluation with MicroBooNE simulation}},\ }\href {https://doi.org/10.1088/1748-0221/13/07/P07006} {\bibfield  {journal} {\bibinfo  {journal} {{J. Instrum.}}\ }\textbf {\bibinfo {volume} {13}},\ \bibinfo {pages} {P07006} (\bibinfo {year} {2018}{\natexlab{a}})}\BibitemShut {NoStop}%
\bibitem [{\citenamefont {Adams}\ \emph {et~al.}(2018{\natexlab{b}})\citenamefont {Adams} \emph {et~al.}}]{MicroBooNE:2018vro}%
  \BibitemOpen
  \bibfield  {author} {\bibinfo {author} {\bibfnamefont {C.}~\bibnamefont {Adams}} \emph {et~al.} (\bibinfo {collaboration} {{MicroBooNE}}),\ }\bibfield  {title} {\bibinfo {title} {{Ionization electron signal processing in single phase LArTPCs. Part II. Data/simulation comparison and performance in MicroBooNE}},\ }\href {https://doi.org/10.1088/1748-0221/13/07/P07007} {\bibfield  {journal} {\bibinfo  {journal} {{J. Instrum.}}\ }\textbf {\bibinfo {volume} {13}},\ \bibinfo {pages} {P07007} (\bibinfo {year} {2018}{\natexlab{b}})}\BibitemShut {NoStop}%
\bibitem [{\citenamefont {Snider}\ and\ \citenamefont {Petrillo}(2017)}]{larsoft}%
  \BibitemOpen
  \bibfield  {author} {\bibinfo {author} {\bibfnamefont {E.}~\bibnamefont {Snider}}\ and\ \bibinfo {author} {\bibfnamefont {G.}~\bibnamefont {Petrillo}},\ }\bibfield  {title} {\bibinfo {title} {{LArSoft: toolkit for simulation, reconstruction and analysis of liquid argon TPC neutrino detectors}},\ }\href {https://doi.org/10.1088/1742-6596/898/4/042057} {\bibfield  {journal} {\bibinfo  {journal} {J. Phys. Conf. Ser.}\ }\textbf {\bibinfo {volume} {898}},\ \bibinfo {pages} {042057} (\bibinfo {year} {2017})}\BibitemShut {NoStop}%
\bibitem [{\citenamefont {Baller}(2017)}]{Baller:2017ugz}%
  \BibitemOpen
  \bibfield  {author} {\bibinfo {author} {\bibfnamefont {B.}~\bibnamefont {Baller}},\ }\bibfield  {title} {\bibinfo {title} {{Liquid argon TPC signal formation, signal processing and reconstruction techniques}},\ }\href {https://doi.org/10.1088/1748-0221/12/07/P07010} {\bibfield  {journal} {\bibinfo  {journal} {{J. Instrum.}}\ }\textbf {\bibinfo {volume} {12}},\ \bibinfo {pages} {P07010} (\bibinfo {year} {2017})}\BibitemShut {NoStop}%
\bibitem [{\citenamefont {Miyajima}\ \emph {et~al.}(1974)\citenamefont {Miyajima} \emph {et~al.}}]{PhysRevA.9.1438}%
  \BibitemOpen
  \bibfield  {author} {\bibinfo {author} {\bibfnamefont {M.}~\bibnamefont {Miyajima}} \emph {et~al.},\ }\bibfield  {title} {\bibinfo {title} {Average energy expended per ion pair in liquid argon},\ }\href {https://doi.org/10.1103/PhysRevA.9.1438} {\bibfield  {journal} {\bibinfo  {journal} {Phys. Rev. A}\ }\textbf {\bibinfo {volume} {9}},\ \bibinfo {pages} {1438} (\bibinfo {year} {1974})}\BibitemShut {NoStop}%
\bibitem [{\citenamefont {Sumner}(2002)}]{DM_Review}%
  \BibitemOpen
  \bibfield  {author} {\bibinfo {author} {\bibfnamefont {T.}~\bibnamefont {Sumner}},\ }\bibfield  {title} {\bibinfo {title} {{Experimental Searches for Dark Matter}},\ }\href {https://doi.org/https://doi.org/10.12942/lrr-2002-4} {\bibfield  {journal} {\bibinfo  {journal} {{Living Rev. Relativ.}}\ }\textbf {\bibinfo {volume} {5}},\ \bibinfo {pages} {4} (\bibinfo {year} {2002})}\BibitemShut {NoStop}%
\bibitem [{\citenamefont {Auger}\ \emph {et~al.}(2012)\citenamefont {Auger} \emph {et~al.}}]{EXO-200:2012pdt}%
  \BibitemOpen
  \bibfield  {author} {\bibinfo {author} {\bibfnamefont {M.}~\bibnamefont {Auger}} \emph {et~al.} (\bibinfo {collaboration} {EXO-200}),\ }\bibfield  {title} {\bibinfo {title} {{Search for Neutrinoless Double-Beta Decay in $^{136}$Xe with EXO-200}},\ }\href {https://doi.org/10.1103/PhysRevLett.109.032505} {\bibfield  {journal} {\bibinfo  {journal} {Phys. Rev. Lett.}\ }\textbf {\bibinfo {volume} {109}},\ \bibinfo {pages} {032505} (\bibinfo {year} {2012})}\BibitemShut {NoStop}%
\bibitem [{\citenamefont {Akerib}\ \emph {et~al.}(2015)\citenamefont {Akerib} \emph {et~al.}}]{LZ:2015kxe}%
  \BibitemOpen
  \bibfield  {author} {\bibinfo {author} {\bibfnamefont {D.~S.}\ \bibnamefont {Akerib}} \emph {et~al.} (\bibinfo {collaboration} {LZ}),\ }\bibfield  {title} {\bibinfo {title} {{LUX-ZEPLIN (LZ) Conceptual Design Report}},\ }\href@noop {} {\  (\bibinfo {year} {2015})},\ \Eprint {https://arxiv.org/abs/1509.02910} {arXiv:1509.02910 [physics.ins-det]} \BibitemShut {NoStop}%
\bibitem [{\citenamefont {Alvis}\ \emph {et~al.}(2019)\citenamefont {Alvis} \emph {et~al.}}]{Majorana:2019ftu}%
  \BibitemOpen
  \bibfield  {author} {\bibinfo {author} {\bibfnamefont {S.~I.}\ \bibnamefont {Alvis}} \emph {et~al.} (\bibinfo {collaboration} {Majorana}),\ }\bibfield  {title} {\bibinfo {title} {{Multisite event discrimination for the Majorana demonstrator}},\ }\href {https://doi.org/10.1103/PhysRevC.99.065501} {\bibfield  {journal} {\bibinfo  {journal} {Phys. Rev. C}\ }\textbf {\bibinfo {volume} {99}},\ \bibinfo {pages} {065501} (\bibinfo {year} {2019})}\BibitemShut {NoStop}%
\bibitem [{\citenamefont {Dunger}\ and\ \citenamefont {Biller}(2019)}]{Dunger:2019dfo}%
  \BibitemOpen
  \bibfield  {author} {\bibinfo {author} {\bibfnamefont {J.}~\bibnamefont {Dunger}}\ and\ \bibinfo {author} {\bibfnamefont {S.~D.}\ \bibnamefont {Biller}},\ }\bibfield  {title} {\bibinfo {title} {{Multi-site Event Discrimination in Large Liquid Scintillation Detectors}},\ }\href {https://doi.org/10.1016/j.nima.2019.162420} {\bibfield  {journal} {\bibinfo  {journal} {Nucl. Instrum. Meth. A}\ }\textbf {\bibinfo {volume} {943}},\ \bibinfo {pages} {162420} (\bibinfo {year} {2019})}\BibitemShut {NoStop}%
\bibitem [{\citenamefont {Fischer}\ \emph {et~al.}(2019)\citenamefont {Fischer} \emph {et~al.}}]{Fischer:2019qfr}%
  \BibitemOpen
  \bibfield  {author} {\bibinfo {author} {\bibfnamefont {V.}~\bibnamefont {Fischer}} \emph {et~al.},\ }\bibfield  {title} {\bibinfo {title} {{Measurement of the neutron capture cross-section on argon}},\ }\href {https://doi.org/10.1103/PhysRevD.99.103021} {\bibfield  {journal} {\bibinfo  {journal} {Phys. Rev. D}\ }\textbf {\bibinfo {volume} {99}},\ \bibinfo {pages} {103021} (\bibinfo {year} {2019})}\BibitemShut {NoStop}%
\bibitem [{\citenamefont {Abratenko}\ \emph {et~al.}(2022{\natexlab{e}})\citenamefont {Abratenko} \emph {et~al.}}]{ub_syst}%
  \BibitemOpen
  \bibfield  {author} {\bibinfo {author} {\bibfnamefont {P.}~\bibnamefont {Abratenko}} \emph {et~al.} (\bibinfo {collaboration} {MicroBooNE}),\ }\bibfield  {title} {\bibinfo {title} {{Novel approach for evaluating detector-related uncertainties in a LArTPC using MicroBooNE data}},\ }\href {https://doi.org/10.1140/epjc/s10052-022-10270-8} {\bibfield  {journal} {\bibinfo  {journal} {Eur. Phys. J. C}\ }\textbf {\bibinfo {volume} {82}},\ \bibinfo {pages} {454} (\bibinfo {year} {2022}{\natexlab{e}})}\BibitemShut {NoStop}%
\bibitem [{\citenamefont {Abratenko}\ \emph {et~al.}(2021{\natexlab{a}})\citenamefont {Abratenko} \emph {et~al.}}]{ub_diffusion:}%
  \BibitemOpen
  \bibfield  {author} {\bibinfo {author} {\bibfnamefont {P.}~\bibnamefont {Abratenko}} \emph {et~al.} (\bibinfo {collaboration} {{MicroBooNE}}),\ }\bibfield  {title} {\bibinfo {title} {{Measurement of the longitudinal diffusion of ionization electrons in the MicroBooNE detector}},\ }\href {https://doi.org/10.1088/1748-0221/16/09/P09025} {\bibfield  {journal} {\bibinfo  {journal} {{J. Instrum.}}\ }\textbf {\bibinfo {volume} {16}},\ \bibinfo {pages} {P09025} (\bibinfo {year} {2021}{\natexlab{a}})}\BibitemShut {NoStop}%
\bibitem [{\citenamefont {Abratenko}\ \emph {et~al.}(2021{\natexlab{b}})\citenamefont {Abratenko} \emph {et~al.}}]{MicroBooNE:2020fmc}%
  \BibitemOpen
  \bibfield  {author} {\bibinfo {author} {\bibfnamefont {P.}~\bibnamefont {Abratenko}} \emph {et~al.} (\bibinfo {collaboration} {MicroBooNE}),\ }\bibfield  {title} {\bibinfo {title} {{Measurement of the atmospheric muon rate with the MicroBooNE Liquid Argon TPC}},\ }\href {https://doi.org/10.1088/1748-0221/16/04/P04004} {\bibfield  {journal} {\bibinfo  {journal} {J. Instrum.}\ }\textbf {\bibinfo {volume} {16}}\bibinfo  {number} { (04)},\ \bibinfo {pages} {P04004}}\BibitemShut {NoStop}%
\bibitem [{\citenamefont {Adhikari}\ \emph {et~al.}(2023)\citenamefont {Adhikari} \emph {et~al.}}]{DEAP:2023wri}%
  \BibitemOpen
\bibfield  {number} {  }\bibfield  {author} {\bibinfo {author} {\bibfnamefont {P.}~\bibnamefont {Adhikari}} \emph {et~al.} (\bibinfo {collaboration} {DEAP}),\ }\bibfield  {title} {\bibinfo {title} {{Precision measurement of the specific activity of $^{39}$Ar in atmospheric argon with the DEAP-3600 detector}},\ }\href {https://doi.org/10.1140/epjc/s10052-023-11678-6} {\bibfield  {journal} {\bibinfo  {journal} {Eur. Phys. J. C}\ }\textbf {\bibinfo {volume} {83}},\ \bibinfo {pages} {642} (\bibinfo {year} {2023})}\BibitemShut {NoStop}%
\bibitem [{\citenamefont {Benetti}\ \emph {et~al.}(2007)\citenamefont {Benetti} \emph {et~al.}}]{WARP:2006nsa}%
  \BibitemOpen
  \bibfield  {author} {\bibinfo {author} {\bibfnamefont {P.}~\bibnamefont {Benetti}} \emph {et~al.} (\bibinfo {collaboration} {WARP}),\ }\bibfield  {title} {\bibinfo {title} {{Measurement of the specific activity of Ar-39 in natural argon}},\ }\href {https://doi.org/10.1016/j.nima.2007.01.106} {\bibfield  {journal} {\bibinfo  {journal} {Nucl. Instrum. Meth. A}\ }\textbf {\bibinfo {volume} {574}},\ \bibinfo {pages} {83} (\bibinfo {year} {2007})}\BibitemShut {NoStop}%
\bibitem [{\citenamefont {{National Nuclear Data Center (NNDC), Brookhaven National Laboratory}}(2024)}]{nndc_chart}%
  \BibitemOpen
  \bibfield  {author} {\bibinfo {author} {\bibnamefont {{National Nuclear Data Center (NNDC), Brookhaven National Laboratory}}},\ }\href {https://www.nndc.bnl.gov/nudat3} {\bibinfo {title} {{NuDat 3.0 Database}}} (\bibinfo {year} {2024})\BibitemShut {NoStop}%
\bibitem [{\citenamefont {Novella}\ \emph {et~al.}(2018)\citenamefont {Novella} \emph {et~al.}}]{NEXT:2018zho}%
  \BibitemOpen
  \bibfield  {author} {\bibinfo {author} {\bibfnamefont {P.}~\bibnamefont {Novella}} \emph {et~al.} (\bibinfo {collaboration} {NEXT}),\ }\bibfield  {title} {\bibinfo {title} {{Measurement of radon-induced backgrounds in the NEXT double beta decay experiment}},\ }\href {https://doi.org/10.1007/JHEP10(2018)112} {\bibfield  {journal} {\bibinfo  {journal} {JHEP}\ }\textbf {\bibinfo {volume} {10}},\ \bibinfo {pages} {112}}\BibitemShut {NoStop}%
\bibitem [{\citenamefont {Adey}\ \emph {et~al.}(2019)\citenamefont {Adey} \emph {et~al.}}]{DayaBay:2019fje}%
  \BibitemOpen
  \bibfield  {author} {\bibinfo {author} {\bibfnamefont {D.}~\bibnamefont {Adey}} \emph {et~al.} (\bibinfo {collaboration} {Daya Bay}),\ }\bibfield  {title} {\bibinfo {title} {{A high precision calibration of the nonlinear energy response at Daya Bay}},\ }\href {https://doi.org/10.1016/j.nima.2019.06.031} {\bibfield  {journal} {\bibinfo  {journal} {Nucl. Instrum. Meth. A}\ }\textbf {\bibinfo {volume} {940}},\ \bibinfo {pages} {230} (\bibinfo {year} {2019})}\BibitemShut {NoStop}%
\bibitem [{\citenamefont {Bellini}\ \emph {et~al.}(2010)\citenamefont {Bellini} \emph {et~al.}}]{Borexino:2009mcw}%
  \BibitemOpen
  \bibfield  {author} {\bibinfo {author} {\bibfnamefont {G.}~\bibnamefont {Bellini}} \emph {et~al.} (\bibinfo {collaboration} {Borexino}),\ }\bibfield  {title} {\bibinfo {title} {{New experimental limits on the Pauli forbidden transitions in C-12 nuclei obtained with 485 days Borexino data}},\ }\href {https://doi.org/10.1103/PhysRevC.81.034317} {\bibfield  {journal} {\bibinfo  {journal} {Phys. Rev. C}\ }\textbf {\bibinfo {volume} {81}},\ \bibinfo {pages} {034317} (\bibinfo {year} {2010})}\BibitemShut {NoStop}%
\bibitem [{\citenamefont {Albert}\ \emph {et~al.}(2014)\citenamefont {Albert} \emph {et~al.}}]{EXO-200:2014ofj}%
  \BibitemOpen
  \bibfield  {author} {\bibinfo {author} {\bibfnamefont {J.~B.}\ \bibnamefont {Albert}} \emph {et~al.} (\bibinfo {collaboration} {EXO-200}),\ }\bibfield  {title} {\bibinfo {title} {{Search for Majorana neutrinos with the first two years of EXO-200 data}},\ }\href {https://doi.org/10.1038/nature13432} {\bibfield  {journal} {\bibinfo  {journal} {Nature}\ }\textbf {\bibinfo {volume} {510}},\ \bibinfo {pages} {229} (\bibinfo {year} {2014})}\BibitemShut {NoStop}%
\bibitem [{\citenamefont {Andriamirado}\ \emph {et~al.}(2021)\citenamefont {Andriamirado} \emph {et~al.}}]{PROSPECT:2020sxr}%
  \BibitemOpen
  \bibfield  {author} {\bibinfo {author} {\bibfnamefont {M.}~\bibnamefont {Andriamirado}} \emph {et~al.} (\bibinfo {collaboration} {PROSPECT}),\ }\bibfield  {title} {\bibinfo {title} {{Improved short-baseline neutrino oscillation search and energy spectrum measurement with the PROSPECT experiment at HFIR}},\ }\href {https://doi.org/10.1103/PhysRevD.103.032001} {\bibfield  {journal} {\bibinfo  {journal} {Phys. Rev. D}\ }\textbf {\bibinfo {volume} {103}},\ \bibinfo {pages} {032001} (\bibinfo {year} {2021})}\BibitemShut {NoStop}%
\bibitem [{\citenamefont {Acciarri}\ \emph {et~al.}(2013)\citenamefont {Acciarri} \emph {et~al.}}]{argo_modbox}%
  \BibitemOpen
  \bibfield  {author} {\bibinfo {author} {\bibfnamefont {R.}~\bibnamefont {Acciarri}} \emph {et~al.} (\bibinfo {collaboration} {ArgoNeuT Collaboration}),\ }\bibfield  {title} {\bibinfo {title} {{A study of electron recombination using highly ionizing particles in the ArgoNeuT Liquid Argon TPC}},\ }\href {https://doi.org/10.1088/1748-0221/8/08/P08005} {\bibfield  {journal} {\bibinfo  {journal} {{J. Instrum.}}\ }\textbf {\bibinfo {volume} {8}},\ \bibinfo {pages} {P08005} (\bibinfo {year} {2013})}\BibitemShut {NoStop}%
\bibitem [{\citenamefont {Adams}\ \emph {et~al.}(2020)\citenamefont {Adams} \emph {et~al.}}]{ub_cal}%
  \BibitemOpen
  \bibfield  {author} {\bibinfo {author} {\bibfnamefont {C.}~\bibnamefont {Adams}} \emph {et~al.} (\bibinfo {collaboration} {MicroBooNE}),\ }\bibfield  {title} {\bibinfo {title} {{Calibration of the charge and energy loss per unit length of the MicroBooNE liquid argon time projection chamber using muons and protons}},\ }\href {https://doi.org/10.1088/1748-0221/15/03/P03022} {\bibfield  {journal} {\bibinfo  {journal} {{J. Instrum.}}\ }\textbf {\bibinfo {volume} {15}},\ \bibinfo {pages} {P03022} (\bibinfo {year} {2020})}\BibitemShut {NoStop}%
\bibitem [{\citenamefont {Abratenko}\ \emph {et~al.}(2020)\citenamefont {Abratenko} \emph {et~al.}}]{Abratenko_2020}%
  \BibitemOpen
  \bibfield  {author} {\bibinfo {author} {\bibfnamefont {P.}~\bibnamefont {Abratenko}} \emph {et~al.} (\bibinfo {collaboration} {{MicroBooNE}}),\ }\bibfield  {title} {\bibinfo {title} {Measurement of space charge effects in the {MicroBooNE} {LArTPC} using cosmic muons},\ }\href {https://doi.org/10.1088/1748-0221/15/12/p12037} {\bibfield  {journal} {\bibinfo  {journal} {{J. Instrum.}}\ }\textbf {\bibinfo {volume} {15}},\ \bibinfo {pages} {P12037} (\bibinfo {year} {2020})}\BibitemShut {NoStop}%
\bibitem [{\citenamefont {{Pacific Northwest National Laboratory and SNOLAB}}(2024)}]{radiopurity}%
  \BibitemOpen
  \bibfield  {author} {\bibinfo {author} {\bibnamefont {{Pacific Northwest National Laboratory and SNOLAB}}},\ }\href {https://www.radiopurity.org/} {\bibinfo {title} {{radiopurity.org}}} (\bibinfo {year} {2024})\BibitemShut {NoStop}%
\bibitem [{\citenamefont {Lawson}\ and\ \citenamefont {Cleveland}(2011)}]{Lawson:2011zz}%
  \BibitemOpen
  \bibfield  {author} {\bibinfo {author} {\bibfnamefont {I.}~\bibnamefont {Lawson}}\ and\ \bibinfo {author} {\bibfnamefont {B.}~\bibnamefont {Cleveland}},\ }\bibfield  {title} {\bibinfo {title} {{Low background counting at SNOLAB}},\ }\href {https://doi.org/10.1063/1.3579561} {\bibfield  {journal} {\bibinfo  {journal} {AIP Conf. Proc.}\ }\textbf {\bibinfo {volume} {1338}},\ \bibinfo {pages} {68} (\bibinfo {year} {2011})}\BibitemShut {NoStop}%
\bibitem [{\citenamefont {Arpesella}\ \emph {et~al.}(2002)\citenamefont {Arpesella} \emph {et~al.}}]{BOREXINO:2001bob}%
  \BibitemOpen
  \bibfield  {author} {\bibinfo {author} {\bibfnamefont {C.}~\bibnamefont {Arpesella}} \emph {et~al.} (\bibinfo {collaboration} {BOREXINO}),\ }\bibfield  {title} {\bibinfo {title} {{Measurements of extremely low radioactivity levels in BOREXINO}},\ }\href {https://doi.org/10.1016/S0927-6505(01)00179-7} {\bibfield  {journal} {\bibinfo  {journal} {Astropart. Phys.}\ }\textbf {\bibinfo {volume} {18}},\ \bibinfo {pages} {1} (\bibinfo {year} {2002})}\BibitemShut {NoStop}%
\bibitem [{\citenamefont {Gardiner}(2021)}]{Gardiner:2021qfr}%
  \BibitemOpen
  \bibfield  {author} {\bibinfo {author} {\bibfnamefont {S.}~\bibnamefont {Gardiner}},\ }\bibfield  {title} {\bibinfo {title} {{Simulating low-energy neutrino interactions with MARLEY}},\ }\href {https://doi.org/10.1016/j.cpc.2021.108123} {\bibfield  {journal} {\bibinfo  {journal} {Comput. Phys. Commun.}\ }\textbf {\bibinfo {volume} {269}},\ \bibinfo {pages} {108123} (\bibinfo {year} {2021})}\BibitemShut {NoStop}%
\bibitem [{\citenamefont {Asaadi}\ \emph {et~al.}(2022)\citenamefont {Asaadi} \emph {et~al.}}]{Asaadi:2022ojm}%
  \BibitemOpen
  \bibfield  {author} {\bibinfo {author} {\bibfnamefont {J.}~\bibnamefont {Asaadi}} \emph {et~al.},\ }\bibfield  {title} {\bibinfo {title} {{Physics Opportunities in the ORNL Spallation Neutron Source Second Target Station Era}},\ }in\ \href@noop {} {\emph {\bibinfo {booktitle} {{Snowmass 2021}}}}\ (\bibinfo {year} {2022})\ \Eprint {https://arxiv.org/abs/2209.02883} {arXiv:2209.02883 [hep-ex]} \BibitemShut {NoStop}%
\bibitem [{\citenamefont {Aguilar-Arevalo}\ \emph {et~al.}(2023)\citenamefont {Aguilar-Arevalo} \emph {et~al.}}]{Aguilar-Arevalo2023dai}%
  \BibitemOpen
  \bibfield  {author} {\bibinfo {author} {\bibfnamefont {A.~A.}\ \bibnamefont {Aguilar-Arevalo}} \emph {et~al.},\ }\bibfield  {title} {\bibinfo {title} {{Physics Opportunities at a Beam Dump Facility at PIP-II at Fermilab and Beyond}},\ }\href@noop {} {\  (\bibinfo {year} {2023})},\ \Eprint {https://arxiv.org/abs/2311.09915} {arXiv:2311.09915 [hep-ex]} \BibitemShut {NoStop}%
\bibitem [{\citenamefont {Bhatia}\ \emph {et~al.}(2012)\citenamefont {Bhatia}, \citenamefont {Finch}, \citenamefont {Gooden},\ and\ \citenamefont {Tornow}}]{PhysRevC.86.041602}%
  \BibitemOpen
  \bibfield  {author} {\bibinfo {author} {\bibfnamefont {C.}~\bibnamefont {Bhatia}}, \bibinfo {author} {\bibfnamefont {S.~W.}\ \bibnamefont {Finch}}, \bibinfo {author} {\bibfnamefont {M.~E.}\ \bibnamefont {Gooden}},\ and\ \bibinfo {author} {\bibfnamefont {W.}~\bibnamefont {Tornow}},\ }\bibfield  {title} {\bibinfo {title} {${}^{40}${A}r($n$,$p$)${}^{40}${C}l reaction cross section between 9 and 15 {M}ev},\ }\href {https://doi.org/10.1103/PhysRevC.86.041602} {\bibfield  {journal} {\bibinfo  {journal} {Phys. Rev. C}\ }\textbf {\bibinfo {volume} {86}},\ \bibinfo {pages} {041602} (\bibinfo {year} {2012})}\BibitemShut {NoStop}%
\bibitem [{\citenamefont {Hagmann}\ \emph {et~al.}(2007)\citenamefont {Hagmann}, \citenamefont {Lange},\ and\ \citenamefont {Wright}}]{Hagmann:2007ziw}%
  \BibitemOpen
  \bibfield  {author} {\bibinfo {author} {\bibfnamefont {C.}~\bibnamefont {Hagmann}}, \bibinfo {author} {\bibfnamefont {D.}~\bibnamefont {Lange}},\ and\ \bibinfo {author} {\bibfnamefont {D.}~\bibnamefont {Wright}},\ }\bibfield  {title} {\bibinfo {title} {Cosmic-ray shower generator ({CRY}) for monte carlo transport codes},\ }in\ \href {https://doi.org/10.1109/NSSMIC.2007.4437209} {\emph {\bibinfo {booktitle} {2007 IEEE Nuclear Science Symposium Conference Record}}},\ Vol.~\bibinfo {volume} {2}\ (\bibinfo {year} {2007})\ pp.\ \bibinfo {pages} {1143--1146}\BibitemShut {NoStop}%
\bibitem [{\citenamefont {{MicroBooNE Collaboration}}(2016)}]{MicroBooNE:2016amq}%
  \BibitemOpen
  \bibfield  {author} {\bibinfo {author} {\bibnamefont {{MicroBooNE Collaboration}}},\ }\bibfield  {title} {\bibinfo {title} {Cosmic shielding studies at microboone},\ }\href {https://www.osti.gov/biblio/1573042} {\bibfield  {journal} {\bibinfo  {journal} {MICROBOONE-NOTE-1005-PUB}\ } (\bibinfo {year} {2016})}\BibitemShut {NoStop}%
\bibitem [{\citenamefont {Bhandari}\ \emph {et~al.}(2019)\citenamefont {Bhandari} \emph {et~al.}}]{CAPTAIN:2019fxo}%
  \BibitemOpen
  \bibfield  {author} {\bibinfo {author} {\bibfnamefont {B.}~\bibnamefont {Bhandari}} \emph {et~al.} (\bibinfo {collaboration} {CAPTAIN}),\ }\bibfield  {title} {\bibinfo {title} {{First Measurement of the Total Neutron Cross Section on Argon Between 100 and 800 MeV}},\ }\href {https://doi.org/10.1103/PhysRevLett.123.042502} {\bibfield  {journal} {\bibinfo  {journal} {Phys. Rev. Lett.}\ }\textbf {\bibinfo {volume} {123}},\ \bibinfo {pages} {042502} (\bibinfo {year} {2019})}\BibitemShut {NoStop}%
\bibitem [{\citenamefont {Martynenko}\ \emph {et~al.}(2023)\citenamefont {Martynenko} \emph {et~al.}}]{CAPTAIN:2022nzf}%
  \BibitemOpen
  \bibfield  {author} {\bibinfo {author} {\bibfnamefont {S.}~\bibnamefont {Martynenko}} \emph {et~al.} (\bibinfo {collaboration} {CAPTAIN}),\ }\bibfield  {title} {\bibinfo {title} {{Measurement of the neutron cross section on argon between 95 and 720~MeV}},\ }\href {https://doi.org/10.1103/PhysRevD.107.072009} {\bibfield  {journal} {\bibinfo  {journal} {Phys. Rev. D}\ }\textbf {\bibinfo {volume} {107}},\ \bibinfo {pages} {072009} (\bibinfo {year} {2023})}\BibitemShut {NoStop}%
\end{thebibliography}%



\begin{widetext}
\textbf{\Large{Supplementary Material}}
\section{General-Purpose MicroBooNE Proton Selection Matrices}
\label{sec:maps}

Other LArTPC experiment analyses or phenomenological studies of neutrino LArTPCs may desire more or less restrictive proton PID cuts than those implemented for the path-finding analysis presented in the main manuscript.  
To provide additional flexibility in the use of MeV-scale PID capabilities demonstrated in this paper, in this appendix we include two-dimensional maps of PID cut rejection factors associated with specific reconstructed \Eb/$ds$ boundaries for MicroBooNE single-particle $e^-$ and $p$ MC datasets.  
To enable use of these maps in studies based on truth-level MC simulations, we also include reconstructed blip response matrices describing the relationship between true energy deposit and reconstructed \Eb for each particle type.  
These matrices, calculated using the output from MicroBooNE-specific MC simulations, are also included as separate data files that accompany this manuscript.

\begin{figure}[tbh]
\includegraphics[width=0.48\textwidth]{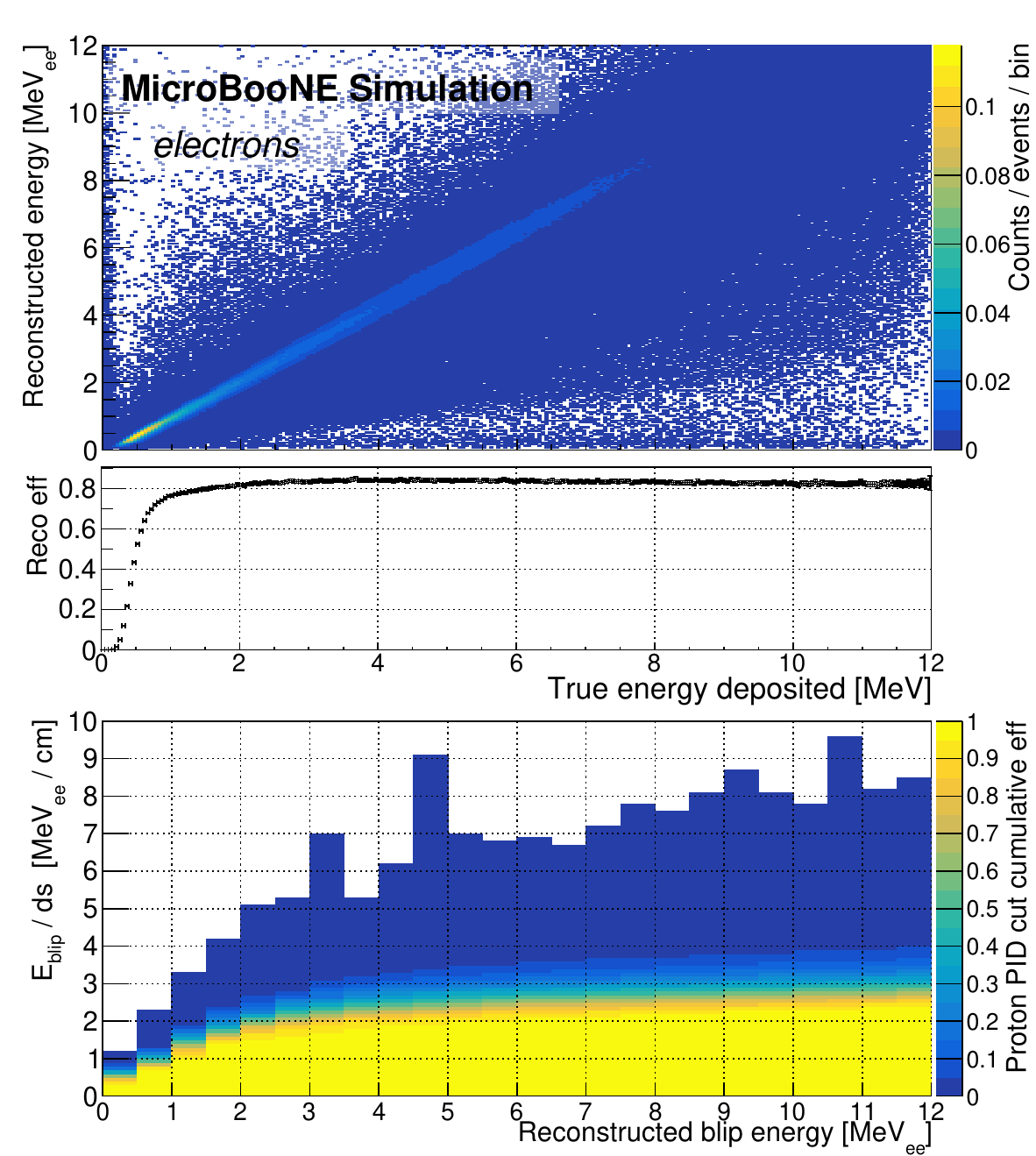}
\includegraphics[width=0.48\textwidth]{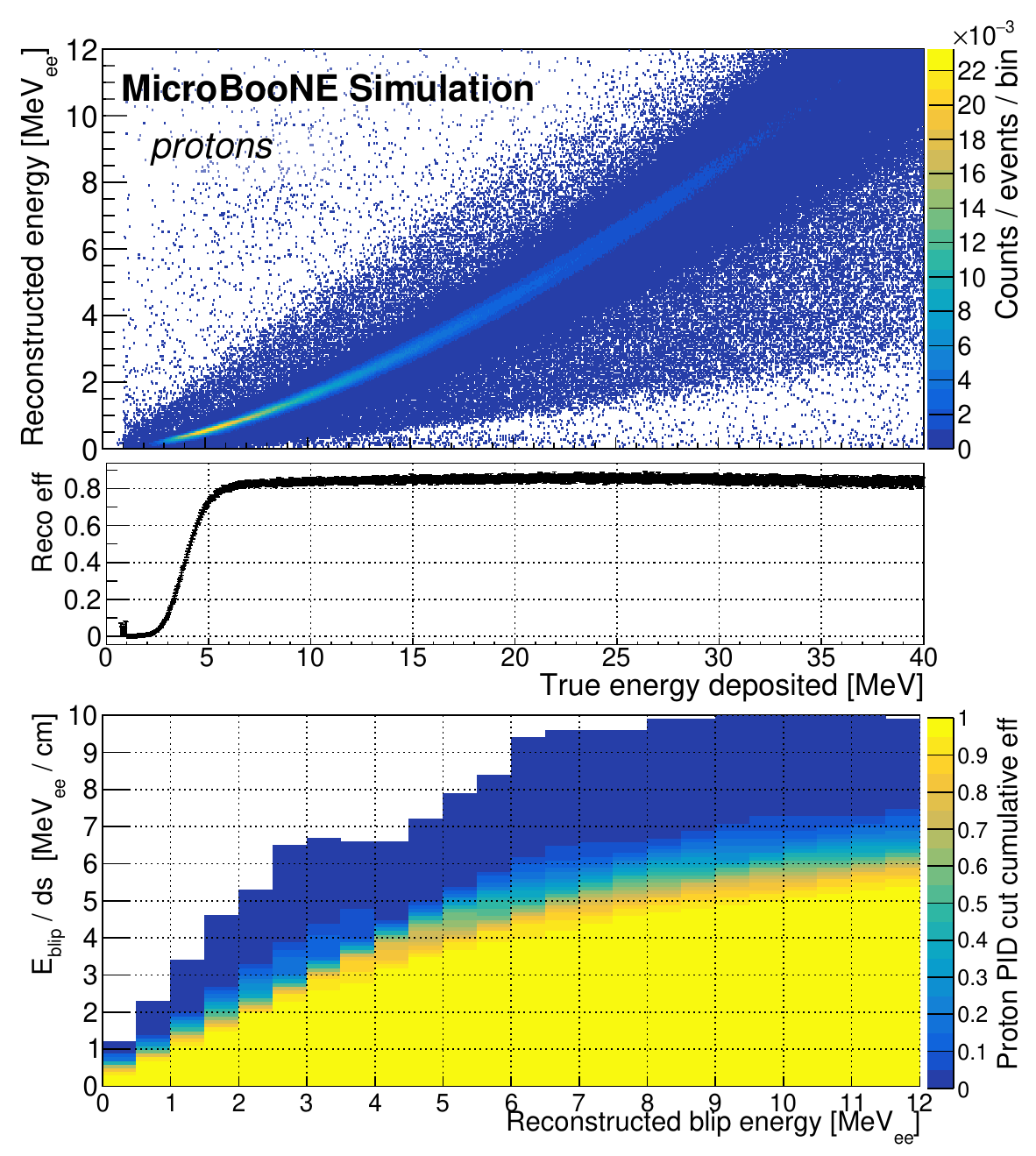}
\caption{Top: MicroBooNE low-energy detector response matrix showing reconstructed blip energy \Eb of simulated uniformly-distributed electrons (left) and protons (right) of varying true energies, prior to application of any PID cuts.  Middle: Total blip detection efficiency delivered by \texttt{BlipReco} in each true energy range in the absence of any PID cuts.  Bottom: Selection sub-efficiency achieved from placement of a PID cut below specific values of \Eb/$ds$ for different \Eb bins for MC electrons and protons generating a reconstructed blip.
}
\label{fig:proton_map}
\end{figure}

Blip response matrices and PID rejection factor maps for the single-particle electron and proton samples are pictured in Fig.~\ref{fig:proton_map}. True energy in these matrices is defined by the true energy deposited in a localized region, as opposed to the true energy of the primary generated particle.  
This definition ensures that true energies are properly assigned for cases where a single particle generates multiple blips -- for example, an electron undergoing radiative energy loss.  
We note that this multi-blip scenario is sub-dominant for the considered proton and electron energy ranges.  
For blip response matrices, the integral of each column corresponds to the total detection efficiency in that true deposited energy range, which is depicted more explicitly in a sub-panel accompanying each matrix.  
Blip detection efficiency reaches a maximum around 85\%, a consequence of the presence of dead wires in the MicroBooNE LArTPC; thus, when applying depicted responses to model another LArTPC's potential capabilities, one may consider correcting for this known MicroBooNE defect. In addition, when modeling the response of a LArTPC with an electric field higher than MicroBooNE’s nominal 274~V/cm, one may also consider modifications to reflect expected improvements in efficiency at the low-energy threshold shoulder due to less charge loss from recombination.

The blip response matrix for electrons generally follows the expected features previously described in Sec.~\ref{sec:reconstruction} of the main manuscript:
blip reconstruction becomes efficient in the 0.2-0.5~MeV range, with fairly diagonal response and an energy resolution around roughly 10\%.  
For protons, blip reconstruction efficiency picks up between roughly 2 and 4~MeV true proton energy, with similar resolution to the electron case. For the low-energy electron sample, we also provide response files for the blip selection applied in this paper, as well as response when requiring only reconstructed signals on the lower-threshold collection plane.

The selection efficiency achieved with varying PID cut boundaries is also given in Fig.~\ref{fig:proton_map}.  
Clear offsets in cut efficiency are visible between proton and electron samples at higher \Eb, a reflection of the $p$-$e^{\pm}$ discrimination demonstrated in Sec.~\ref{sec:cosmic}. 
Interestingly, modest offsets are also visible in lower \Eb bins below that considered for the pure cosmogenically-produced blip samples in Sec.~\ref{sec:cosmic}.
Below 1.5~MeV \Eb, where reconstructed size-energy ratios rise linearly in \Eb, PID cut efficiencies are similar for both particle species.  
Above this reconstructed energy, electron \Eb/$ds$ begins to flatten as the average electron's linear travel distance begins to substantially exceed the LArTPC's minimum position resolution.  
This inflection may enable some degree of PID capability down to 10-15~MeV in true proton energy.  
As an example, for protons with \Eb between 2.5 and 3.0~MeV (roughly 13-15 MeV in proton energy according to the top panel of Fig.~\ref{fig:proton_map}), a PID cut placed at 2.9 MeV/cm will deliver 30\% efficiency, while rejecting $>$95\% of electrons reconstructed in the same energy range.

These \texttt{BlipReco} and PID cut response matrices are intended to be useful for LArTPC non-experts to derive realistic low-energy LArTPC signal expectations from truth-level inputs.  
Predicted true energy distributions for protons and electrons generated by a particular physics model can first be cast through the top matrices in Fig.~\ref{fig:proton_map} (also available in the accompanying data files) to obtain a reconstructed blip energy spectrum that accounts for the variation and cut-off in blip reconstruction efficiency with decreasing true energy.  
For studies involving a single final-state particle type, this response matrix operation, which propagates MicroBooNE detector response and \texttt{BlipReco} reconstruction features at low particle energy, is all that is required for signal estimates.  
If a study is concerned with mixed final states, the PID cut sub-efficiencies depicted in Fig.~\ref{fig:proton_map} can also be applied to preferentially select $p$ or $e^{\pm}$ sub-sets, with the user determining the level of stringency of the applied cut for each bin in reconstructed \Eb.

\end{widetext}


\end{document}